\input harvmac
\input epsf

%
%
%
%
\def\unredoffs{} \def\redoffs{\voffset=-.40truein\hoffset=-.40truein}
\def\speclscape{}
%
%
%
%
\newbox\leftpage \newdimen\fullhsize \newdimen\hstitle \newdimen\hsbody
\tolerance=1000\hfuzz=2pt
\catcode`\@=11 
\def\bigans{b }
\def\answ{b }
\ifx\answ\bigans\message{(This will come out unreduced.}
\magnification=1200\unredoffs\baselineskip=16pt plus 2pt minus 1pt
\hsbody=\hsize \hstitle=\hsize 
\else\message{(This will be reduced.} \let\l@r=L
\magnification=1000\baselineskip=16pt plus 2pt minus 1pt
\vsize=7truein \redoffs
\hstitle=8truein\hsbody=4.75truein\fullhsize=10truein\hsize=\hsbody
\output={\ifnum\pageno=0 
    \shipout\vbox{\speclscape{\hsize\fullhsize\makeheadline}
      \hbox to \fullhsize{\hfill\pagebody\hfill}}\advancepageno
    \else
    \almostshipout{\leftline{\vbox{\pagebody\makefootline}}}\advancepageno
    \fi}
\def\almostshipout#1{\if L\l@r \count1=1 \message{[\the\count0.\the\count1]}
        \global\setbox\leftpage=#1 \global\let\l@r=R
   \else \count1=2
    \shipout\vbox{\speclscape{\hsize\fullhsize\makeheadline}
        \hbox to\fullhsize{\box\leftpage\hfil#1}}  \global\let\l@r=L\fi}
\fi
%
\newcount\yearltd\yearltd=\year\advance\yearltd by -1900

\def\Title#1#2{\nopagenumbers\abstractfont\hsize=\hstitle\rightline{#1}%
\vskip 1in\centerline{\titlefont #2}\abstractfont\vskip
.5in\pageno=0}
%
%

\def\draftmode{\message{ DRAFTMODE }\def\draftdate{{\rm preliminary draft:
\number\month/\number\day/\number\yearltd\ \ \hourmin}}%
\headline={\hfil\draftdate}\writelabels\baselineskip=20pt plus 2pt
minus 2pt
   {\count255=\time\divide\count255 by 60 \xdef\hourmin{\number\count255}
    \multiply\count255 by-60\advance\count255 by\time
    \xdef\hourmin{\hourmin:\ifnum\count255<10 0\fi\the\count255}}}
\def\nolabels{\def\wrlabeL##1{}\def\eqlabeL##1{}\def\reflabeL##1{}}
\def\writelabels{\def\wrlabeL##1{\leavevmode\vadjust{\rlap{\smash%
{\line{{\escapechar=` \hfill\rlap{\sevenrm\hskip.03in\string##1}}}}}}}%
\def\eqlabeL##1{{\escapechar-1\rlap{\sevenrm\hskip.05in\string##1}}}%
\def\reflabeL##1{\noexpand\llap{\noexpand\sevenrm\string\string\string##1}}}
\nolabels
%
\global\newcount\secno \global\secno=0 \global\newcount\meqno
\global\meqno=1
\def\newsec#1{\global\advance\secno by1\message{(\the\secno. #1)}
\global\subsecno=0\eqnres@t\noindent{\bf\the\secno. #1}
\writetoca{{\secsym} {#1}}\par\nobreak\medskip\nobreak}
\def\eqnres@t{\xdef\secsym{\the\secno.}\global\meqno=1\bigbreak\bigskip}
\def\sequentialequations{\def\eqnres@t{\bigbreak}}\xdef\secsym{}
\global\newcount\subsecno \global\subsecno=0
\def\subsec#1{\global\advance\subsecno by1\message{(\secsym\the\subsecno. #1)}
\ifnum\lastpenalty>9000\else\bigbreak\fi
\noindent{\it\secsym\the\subsecno. #1}\writetoca{\string\quad
{\secsym\the\subsecno.} {#1}}\par\nobreak\medskip\nobreak}
\def\appendix#1#2{\global\meqno=1\global\subsecno=0\xdef\secsym{\hbox{#1.}}
\bigbreak\bigskip\noindent{\bf Appendix #1. #2}\message{(#1. #2)}
\writetoca{Appendix {#1.} {#2}}\par\nobreak\medskip\nobreak}
%
%
\def\eqnn#1{\xdef #1{(\secsym\the\meqno)}\writedef{#1\leftbracket#1}%
\global\advance\meqno by1\wrlabeL#1}
\def\eqna#1{\xdef #1##1{\hbox{$(\secsym\the\meqno##1)$}}
\writedef{#1\numbersign1\leftbracket#1{\numbersign1}}%
\global\advance\meqno by1\wrlabeL{#1$\{\}$}}
\def\eqn#1#2{\xdef #1{(\secsym\the\meqno)}\writedef{#1\leftbracket#1}%
\global\advance\meqno by1$$#2\eqno#1\eqlabeL#1$$}
%
\newskip\footskip\footskip14pt plus 1pt minus 1pt 
\def\footnotefont{\ninepoint}\def\f@t#1{\footnotefont #1\@foot}
\def\f@@t{\baselineskip\footskip\bgroup\footnotefont\aftergroup\@foot\let\next}
\setbox\strutbox=\hbox{\vrule height9.5pt depth4.5pt width0pt}
\global\newcount\ftno \global\ftno=0
\def\foot{\global\advance\ftno by1\footnote{$^{\the\ftno}$}}
%
\newwrite\ftfile
\def\footend{\def\foot{\global\advance\ftno by1\chardef\wfile=\ftfile
$^{\the\ftno}$\ifnum\ftno=1\immediate\openout\ftfile=foots.tmp\fi%
\immediate\write\ftfile{\noexpand\smallskip%
\noexpand\item{f\the\ftno:\ }\pctsign}\findarg}%
\def\footatend{\vfill\eject\immediate\closeout\ftfile{\parindent=20pt
\centerline{\bf Footnotes}\nobreak\bigskip\input foots.tmp }}}
\def\footatend{}
%
%
\global\newcount\refno \global\refno=1
\newwrite\rfile
\def\ref{[\the\refno]\nref}
\def\nref#1{\xdef#1{[\the\refno]}\writedef{#1\leftbracket#1}%
\ifnum\refno=1\immediate\openout\rfile=refs.tmp\fi
\global\advance\refno by1\chardef\wfile=\rfile\immediate
\write\rfile{\noexpand\item{#1\
}\reflabeL{#1\hskip.31in}\pctsign}\findarg}
\def\findarg#1#{\begingroup\obeylines\newlinechar=`\^^M\pass@rg}
{\obeylines\gdef\pass@rg#1{\writ@line\relax #1^^M\hbox{}^^M}%
\gdef\writ@line#1^^M{\expandafter\toks0\expandafter{\striprel@x #1}%
\edef\next{\the\toks0}\ifx\next\em@rk\let\next=\endgroup\else\ifx\next\empty%
\else\immediate\write\wfile{\the\toks0}\fi\let\next=\writ@line\fi\next\relax}}
\def\striprel@x#1{} \def\em@rk{\hbox{}}
\def\lref{\begingroup\obeylines\lr@f}
\def\lr@f#1#2{\gdef#1{\ref#1{#2}}\endgroup\unskip}

\def\addref#1{\immediate\write\rfile{\noexpand\item{}#1}} 
\def\startrefs#1{\immediate\openout\rfile=refs.tmp\refno=#1}
\def\xref{\expandafter\xr@f}\def\xr@f[#1]{#1}
\def\refs#1{\count255=1[\r@fs #1{\hbox{}}]}
\def\r@fs#1{\ifx\und@fined#1\message{reflabel \string#1 is undefined.}%
\nref#1{need to supply reference \string#1.}\fi%
\vphantom{\hphantom{#1}}\edef\next{#1}\ifx\next\em@rk\def\next{}%
\else\ifx\next#1\ifodd\count255\relax\xref#1\count255=0\fi%
\else#1\count255=1\fi\let\next=\r@fs\fi\next}
%

%
\newwrite\ffile\global\newcount\figno \global\figno=1
\def\fig{fig.~\the\figno\nfig}
\def\nfig#1{\xdef#1{fig.~\the\figno}%
\writedef{#1\leftbracket fig.\noexpand~\the\figno}%
\ifnum\figno=1\immediate\openout\ffile=figs.tmp\fi\chardef\wfile=\ffile%
\immediate\write\ffile{\noexpand\medskip\noexpand\item{Fig.\
\the\figno. }
\reflabeL{#1\hskip.55in}\pctsign}\global\advance\figno
by1\findarg}
\def\xfig{\expandafter\xf@g}\def\xf@g fig.\penalty\@M\ {}
\def\figs#1{figs.~\f@gs #1{\hbox{}}}
\def\f@gs#1{\edef\next{#1}\ifx\next\em@rk\def\next{}\else
\ifx\next#1\xfig #1\else#1\fi\let\next=\f@gs\fi\next}
\newwrite\lfile
{\escapechar-1\xdef\pctsign{\string\%}\xdef\leftbracket{\string\{}
\xdef\rightbracket{\string\}}\xdef\numbersign{\string\#}}

\def\writestop{\def\writestoppt{\immediate\write\lfile{\string\pageno%
\the\pageno\string\startrefs\leftbracket\the\refno\rightbracket%
\string\def\string\secsym\leftbracket\secsym\rightbracket%
\string\secno\the\secno\string\meqno\the\meqno}\immediate\closeout\lfile}}
\def\writestoppt{}\def\writedef#1{}
\def\seclab#1{\xdef #1{\the\secno}\writedef{#1\leftbracket#1}\wrlabeL{#1=#1}}
\def\subseclab#1{\xdef #1{\secsym\the\subsecno}%
\writedef{#1\leftbracket#1}\wrlabeL{#1=#1}}
\newwrite\tfile \def\writetoca#1{}
\def\leaderfill{\leaders\hbox to 1em{\hss.\hss}\hfill}
\def\writetoc{\immediate\openout\tfile=toc.tmp
     \def\writetoca##1{{\edef\next{\write\tfile{\noindent ##1
     \string\leaderfill {\noexpand\number\pageno} \par}}\next}}}

%
\catcode`\@=12 
%
\edef\tfontsize{\ifx\answ\bigans scaled\magstep3\else
scaled\magstep4\fi} \font\titlerm=cmr10 \tfontsize
\font\titlerms=cmr7 \tfontsize \font\titlermss=cmr5 \tfontsize
\font\titlei=cmmi10 \tfontsize \font\titleis=cmmi7 \tfontsize
\font\titleiss=cmmi5 \tfontsize \font\titlesy=cmsy10 \tfontsize
\font\titlesys=cmsy7 \tfontsize \font\titlesyss=cmsy5 \tfontsize
\font\titleit=cmti10 \tfontsize \skewchar\titlei='177
\skewchar\titleis='177 \skewchar\titleiss='177
\skewchar\titlesy='60 \skewchar\titlesys='60
\skewchar\titlesyss='60
\def\titlefont{\def\rm{\fam0\titlerm}
\textfont0=\titlerm \scriptfont0=\titlerms
\scriptscriptfont0=\titlermss \textfont1=\titlei
\scriptfont1=\titleis \scriptscriptfont1=\titleiss
\textfont2=\titlesy \scriptfont2=\titlesys
\scriptscriptfont2=\titlesyss \textfont\itfam=\titleit
\def\it{\fam\itfam\titleit}\rm}
 \ifx\answ\bigans\else scaled\magstep1\fi
\ifx\answ\bigans\def\abstractfont{\tenpoint}\else
\font\abssl=cmsl10 scaled \magstep1 \font\absrm=cmr10
scaled\magstep1 \font\absrms=cmr7 scaled\magstep1
\font\absrmss=cmr5 scaled\magstep1 \font\absi=cmmi10
scaled\magstep1 \font\absis=cmmi7 scaled\magstep1
\font\absiss=cmmi5 scaled\magstep1 \font\abssy=cmsy10
scaled\magstep1 \font\abssys=cmsy7 scaled\magstep1
\font\abssyss=cmsy5 scaled\magstep1 \font\absbf=cmbx10
scaled\magstep1 \skewchar\absi='177 \skewchar\absis='177
\skewchar\absiss='177 \skewchar\abssy='60 \skewchar\abssys='60
\skewchar\abssyss='60
\def\abstractfont{\def\rm{\fam0\absrm}
\textfont0=\absrm \scriptfont0=\absrms \scriptscriptfont0=\absrmss
\textfont1=\absi \scriptfont1=\absis \scriptscriptfont1=\absiss
\textfont2=\abssy \scriptfont2=\abssys \scriptscriptfont2=\abssyss
\textfont\itfam=\bigit \def\it{\fam\itfam\bigit}\def\footnotefont{\tenpoint}%
\textfont\slfam=\abssl \def\sl{\fam\slfam\abssl}%
\textfont\bffam=\absbf \def\bf{\fam\bffam\absbf}\rm}\fi
\def\tenpoint{\def\rm{\fam0\tenrm}
\textfont0=\tenrm \scriptfont0=\sevenrm \scriptscriptfont0=\fiverm
\textfont1=\teni  \scriptfont1=\seveni  \scriptscriptfont1=\fivei
\textfont2=\tensy \scriptfont2=\sevensy \scriptscriptfont2=\fivesy
\textfont\itfam=\tenit \def\it{\fam\itfam\tenit}\def\footnotefont{\ninepoint}%
\textfont\bffam=\tenbf
\def\bf{\fam\bffam\tenbf}\def\sl{\fam\slfam\tensl}\rm}
\font\ninerm=cmr9 \font\sixrm=cmr6 \font\ninei=cmmi9
\font\sixi=cmmi6 \font\ninesy=cmsy9 \font\sixsy=cmsy6
\font\ninebf=cmbx9 \font\nineit=cmti9 \font\ninesl=cmsl9
\skewchar\ninei='177 \skewchar\sixi='177 \skewchar\ninesy='60
\skewchar\sixsy='60
\def\ninepoint{\def\rm{\fam0\ninerm}
\textfont0=\ninerm \scriptfont0=\sixrm \scriptscriptfont0=\fiverm
\textfont1=\ninei \scriptfont1=\sixi \scriptscriptfont1=\fivei
\textfont2=\ninesy \scriptfont2=\sixsy \scriptscriptfont2=\fivesy
\textfont\itfam=\ninei \def\it{\fam\itfam\nineit}\def\sl{\fam\slfam\ninesl}%
\textfont\bffam=\ninebf \def\bf{\fam\bffam\ninebf}\rm}
%
%

\hyphenation{anom-aly anom-alies coun-ter-term coun-ter-terms}
\def\inv{^{\raise.15ex\hbox{${\scriptscriptstyle -}$}\kern-.05em 1}}

\def\Dsl{\,\raise.15ex\hbox{/}\mkern-13.5mu D} 
\def\dsl{\raise.15ex\hbox{/}\kern-.57em\partial}

\font\bigit=cmti10 scaled \magstep1
\def\lspace{\ifx\answ\bigans{}\else\qquad\fi}
\def\lbspace{\ifx\answ\bigans{}\else\hskip-.2in\fi} 
\def\boxeqn#1{\vcenter{\vbox{\hrule\hbox{\vrule\kern3pt\vbox{\kern3pt
      \hbox{${\displaystyle #1}$}\kern3pt}\kern3pt\vrule}\hrule}}}
\def\mbox#1#2{\vcenter{\hrule \hbox{\vrule height#2in
          \kern#1in \vrule} \hrule}}  
%

\def\darr#1{\raise1.5ex\hbox{$\leftrightarrow$}\mkern-16.5mu #1}

\def\roughly#1{\raise.3ex\hbox{$#1$\kern-.75em\lower1ex\hbox{$\sim$}}}

\def\IB{\relax\hbox{$\inbar\kern-.3em{\rm B}$}}

\def\ID{\relax\hbox{$\inbar\kern-.3em{\rm D}$}}
\def\IE{\relax\hbox{$\inbar\kern-.3em{\rm E}$}}
\def\IF{\relax\hbox{$\inbar\kern-.3em{\rm F}$}}
\def\IG{\relax\hbox{$\inbar\kern-.3em{\rm G}$}}
\def\IGa{\relax\hbox{${\rm I}\kern-.18em\Gamma$}}
\def\IH{\relax{\rm I\kern-.18em H}}
\def\IK{\relax{\rm I\kern-.18em K}}
\def\IL{\relax{\rm I\kern-.18em L}}
\def\IP{\relax{\rm I\kern-.18em P}}
\def\IR{{\bf R}}

\def\II{\relax{\rm I\kern-.18em I}}






\def\inbar{\,\vrule height1.5ex width.4pt depth0pt}


\def\kk{{\kappa}}
\def\lref{\begingroup\obeylines\lr@f}
\def\lr@f#1#2{\gdef#1{\ref#1{#2}}\endgroup\unskip}

\lref\ama{
D.~Amati and C.~Klimcik,
``Strings In A Shock Wave Background And Generation
 Of Curved Geometry From Flat Space String Theory,''
Phys.\ Lett.\ B {\bf 210}, 92 (1988).
}

\lref\AmatiWQ{
D.~Amati, M.~Ciafaloni and G.~Veneziano,
``Superstring Collisions At Planckian Energies,''
Phys.\ Lett.\ B {\bf 197}, 81 (1987)~; 
``Classical And Quantum Gravity Effects From Planckian
 Energy Superstring Collisions,''
Int.\ J.\ Mod.\ Phys.\ A {\bf 3}, 1615 (1988).
}

\lref\deV{
H.~J.~de Vega and N.~Sanchez,
``Quantum String Propagation Through Gravitational Shock Waves,''
Phys.\ Lett.\ B {\bf 244}, 215 (1990)~;
``Quantum String Scattering In The Aichelburg-Sexl Geometry,''
Nucl.\ Phys.\ B {\bf 317}, 706 (1989).
}

\lref\steif{
G.~T.~Horowitz and A.~R.~Steif,
``Space-Time Singularities In String Theory,''
Phys.\ Rev.\ Lett.\  {\bf 64}, 260 (1990)~; 
``Strings In Strong Gravitational Fields,''
Phys.\ Rev.\ D {\bf 42}, 1950 (1990).
}

\lref\kk{
E.~Kiritsis and C.~Kounnas,
``String Propagation In Gravitational Wave Backgrounds,''
Phys.\ Lett.\ B {\bf 320}, 264 (1994)
[Addendum-ibid.\ B {\bf 325}, 536 (1994)]
[arXiv:hep-th/9310202]~; 
E.~Kiritsis, C.~Kounnas and D.~Lust,
``Superstring gravitational wave 
backgrounds with space-time supersymmetry,''
Phys.\ Lett.\ B {\bf 331}, 321 (1994)
[arXiv:hep-th/9404114].
}

\lref\met{
R.~R.~Metsaev,
``Type IIB Green-Schwarz superstring
 in plane wave Ramond-Ramond  background,''
Nucl.\ Phys.\ B {\bf 625}, 70 (2002)
[arXiv:hep-th/0112044].
}

\lref\bmn{
D.~Berenstein, J.~M.~Maldacena and H.~Nastase,
``Strings in flat space and pp waves from N = 4 super Yang Mills,''
JHEP {\bf 0204}, 013 (2002)
[arXiv:hep-th/0202021].
}

\lref\mett{
R.~R.~Metsaev and A.~A.~Tseytlin,
``Exactly solvable model of superstring
 in plane wave Ramond-Ramond  background,''
Phys.\ Rev.\ D {\bf 65}, 126004 (2002)
[arXiv:hep-th/0202109].
}

\lref\atish{
A.~Dabholkar and S.~Parvizi,
``Dp branes in pp-wave background,''
Nucl.\ Phys.\ B {\bf 641}, 223 (2002)
[arXiv:hep-th/0203231].
}

\lref\kirp{
E.~Kiritsis and B.~Pioline,
``Strings in homogeneous gravitational waves and null holography,''
JHEP {\bf 0208}, 048 (2002)
[arXiv:hep-th/0204004].
}

\lref\lms{
H.~Liu, G.~Moore and N.~Seiberg,
``Strings in a time-dependent orbifold,''
JHEP {\bf 0206}, 045 (2002)
[arXiv:hep-th/0204168]~; 
``Strings in time-dependent orbifolds,''
JHEP {\bf 0210}, 031 (2002)
[arXiv:hep-th/0206182].
}

\lref\kumar{
A.~Kumar, R.~R.~Nayak and Sanjay,
``D-brane solutions in pp-wave background,''
Phys.\ Lett.\ B {\bf 541}, 183 (2002)
[arXiv:hep-th/0204025].
}

\lref\ske{
K.~Skenderis and M.~Taylor,
``Branes in AdS and pp-wave spacetimes,''
JHEP {\bf 0206}, 025 (2002)
[arXiv:hep-th/0204054]~; 
``Open strings in the plane wave background.
 I: Quantization and  symmetries,''
arXiv:hep-th/0211011.
}

\lref\bgg{
O.~Bergman, M.~R.~Gaberdiel and M.~B.~Green,
``D-brane interactions in type IIB plane-wave background,''
arXiv:hep-th/0205183. 
}

\lref\gg{ M.~R.~Gaberdiel and M.~B.~Green,
``The D-instanton and other supersymmetric
 D-branes in IIB plane-wave  string theory,''
arXiv:hep-th/0211122.
}

\lref\Fab{
M.~Fabinger and J.~McGreevy,
``On Smooth Time-Dependent Orbifolds and Null Singularities,''
arXiv:hep-th/0206196.
}

\lref\maoz{
J.~Maldacena and L.~Maoz,
``Strings on pp-waves and massive two dimensional field theories,''
arXiv:hep-th/0207284.
}

\lref\gava{
D.~Berenstein, E.~Gava, J.~M.~Maldacena, K.~S.~Narain and H.~Nastase,
``Open strings on plane waves and their Yang-Mills duals,''
arXiv:hep-th/0203249.
}

\lref\fuji{
H.~Fuji, K.~Ito and Y.~Sekino,
``Penrose limit and string theories on various brane backgrounds,''
JHEP {\bf 0211}, 005 (2002)
[arXiv:hep-th/0209004].
}

\lref\bh{
C.~Bachas and C.~Hull,
``Null brane intersections,''
arXiv:hep-th/0210269.
}

\lref\mye{
R.~C.~Myers and D.~J.~Winters,
``From D - anti-D pairs to branes in motion,''
arXiv:hep-th/0211042.
}

\lref\papart{
G.~Papadopoulos, J.~G.~Russo and A.~A.~Tseytlin,
``Solvable model of strings in a time-dependent plane-wave background,''
arXiv:hep-th/0211289.
}

\lref\bain{
P.~Bain, K.~Peeters and M.~Zamaklar,
``D-branes in a plane wave from covariant open strings,''
arXiv:hep-th/0208038.
}

\lref\Oku{
K.~Okuyama,
``D-branes on the null-brane,''
arXiv:hep-th/0211218.
}


\lref\rev{C.~P.~Bachas,
``Lectures on D-branes,''
arXiv:hep-th/9806199.
}

\lref\bbg{
C.~P.~Bachas, P.~Bain and M.~B.~Green,
``Curvature terms in D-brane actions and their M-theory origin,''
JHEP {\bf 9905}, 011 (1999)
[arXiv:hep-th/9903210].
}

\lref\fot{
A.~Fotopoulos,
``On (alpha')**2 corrections to the D-brane action for non-geodesic
  world-volume embeddings,''
JHEP {\bf 0109}, 005 (2001)
[arXiv:hep-th/0104146].
}

\lref\TseytlinBH{
A.~A.~Tseytlin,
Nucl.\ Phys.\ B {\bf 475}, 149 (1996)
[arXiv:hep-th/9604035].
}

\lref\tse{ .
}

\lref\cho{
J.~H.~Cho and P.~Oh,
``Super D-helix,''
Phys.\ Rev.\ D {\bf 64}, 106010 (2001)
[arXiv:hep-th/0105095].
}

\lref\mat{D.~Mateos and P.~K.~Townsend,
Phys.\ Rev.\ Lett.\  {\bf 87}, 011602 (2001)
[arXiv:hep-th/0103030]; 
D.~Mateos, S.~Ng and P.~K.~Townsend,
``Supercurves,''
Phys.\ Lett.\ B {\bf 538}, 366 (2002)
[arXiv:hep-th/0204062].
}

\lref\har{
A.~Dabholkar, J.~P.~Gauntlett, J.~A.~Harvey and D.~Waldram,
``Strings as Solitons and  Black Holes as Strings,''
Nucl.\ Phys.\ B {\bf 474}, 85 (1996)
[arXiv:hep-th/9511053].
}

\lref\lun{
O.~Lunin and S.~D.~Mathur,
``Metric of the multiply wound rotating string,''
Nucl.\ Phys.\ B {\bf 610}, 49 (2001)
[arXiv:hep-th/0105136].
}

\lref\TamaryanDA{
S.~Tamaryan, D.~K.~Park and H.~J.~Muller-Kirsten,
arXiv:hep-th/0209239.
}
\lref\hya{
Y.~Hyakutake and N.~Ohta,
``Supertubes and supercurves from M-ribbons,''
Phys.\ Lett.\ B {\bf 539}, 153 (2002)
[arXiv:hep-th/0204161].
}

\lref\zel{Ya.~B.~Zel'dovich and A.~G.~Polnarev, Astron.\ Zh.\
{\bf 51}, 30 (1974) [SoV.\ Astron.\ {\bf 18}, 17 (1974)];
V.~B.~Braginsky and L.~P.~Grishchuk, Zh.\ Eksp.\
Teor.\ Fiz.\ {\bf 89}, 744 (1985) [Sov.\ Phys.\ JETP\ {\bf 62},
427 (1986)]; V.~B.~Braginsky and K.~S.~Thorne, Nature (London)
{\bf 327}, 123 (1987).
}

\lref\as{P.~C.~Aischelburg and R.~U.~Sexl, Gen.\ Rev.\ Grav.\ {\bf 2},
303 (1971). 
}

\lref\thorne{K.~S.~Thorne, ``Gravitational Radiation: 
A New Window Onto the Universe,'' arXiv:gr-qc/9704042.
}

\lref\thooft{
G.~'t Hooft,
``Graviton Dominance In Ultrahigh-Energy Scattering,''
Phys.\ Lett.\ B {\bf 198}, 61 (1987).
}

\lref\chri{
D.~Christodoulou,
``Nonlinear Nature Of Gravitation And Gravitational Wave Experiments,''
Phys.\ Rev.\ Lett.\  {\bf 67}, 1486 (1991).
}

\lref\iz{C.~Itzykson and J.-B.~Zuber, ``Quantum Field Theory,''
McGraw-Hill (1980).
}

\lref\cosmic{T.~K.~Gaisser and T.~Stanev,
``Cosmic Rays,''
Phys.\ Rev. D  {\bf 66}, 010001 (2002).
}

\lref\bpio{ 
C.~Bachas and B.~Pioline,
``High-energy scattering on distant branes,''
JHEP {\bf 9912}, 004 (1999)
[arXiv:hep-th/9909171].
}


\lref\blau{
M.~Blau, J.~Figueroa-O'Farrill, C.~Hull and G.~Papadopoulos,
``A new maximally supersymmetric background of IIB superstring theory,''
JHEP {\bf 0201}, 047 (2002)
[arXiv:hep-th/0110242]~; 
``Penrose limits and maximal supersymmetry,''
Class.\ Quant.\ Grav.\  {\bf 19}, L87 (2002)
[arXiv:hep-th/0201081].
}

\lref\lasers{ P.~W.~ Milonni and J.~H.~ Eberly,
``Lasers'', Wiley, 1988~. 
}

\lref\grif{ J.~B.~Griffiths, ``Colliding Plane Waves in General
Relativity,'' Oxford University Press, 1991~.
}

\lref\elo{
A.~Abouelsaood, C.~G.~Callan, C.~R.~Nappi and S.~A.~Yost,
``Open Strings In Background Gauge Fields,''
Nucl.\ Phys.\ B {\bf 280}, 599 (1987).
}

\lref\eltwo{C.~Bachas and M.~Porrati,
``Pair Creation Of Open Strings In An Electric Field,''
Phys.\ Lett.\ B {\bf 296}, 77 (1992)
[arXiv:hep-th/9209032].
}

\lref\elt{
J.~Ambjorn, Y.~M.~Makeenko, G.~W.~Semenoff and R.~J.~Szabo,
``String theory in electromagnetic fields,''
arXiv:hep-th/0012092.
}

\lref\vil{A.~Vilenkin and E.~P.~S.~Shellard, ``Cosmic Strings
and Other Topological defects,'' Cambridge University Press 1994.
}

\lref\dam{
T.~Damour and A.~Vilenkin,
``Gravitational wave bursts from cosmic strings,''
Phys.\ Rev.\ Lett.\  {\bf 85}, 3761 (2000)
[arXiv:gr-qc/0004075]~; 
``Gravitational wave bursts from cusps and kinks on cosmic strings,''
Phys.\ Rev.\ D {\bf 64}, 064008 (2001)
[arXiv:gr-qc/0104026].
}

\lref\acpeak{
G.~R.~Vincent, M.~Hindmarsh and M.~Sakellariadou,
``Correlations in cosmic string networks,''
Phys.\ Rev.\ D {\bf 55}, 573 (1997)
[arXiv:astro-ph/9606137].
}

\lref\acpeaks{
J.~Magueijo, A.~Albrecht, P.~Ferreira and D.~Coulson,
``The structure of Doppler peaks induced by active perturbations,''
Phys.\ Rev.\ D {\bf 54}, 3727 (1996)
[arXiv:astro-ph/9605047].
}

\lref\acpeakss{
F.~R.~Bouchet, P.~Peter, A.~Riazuelo and M.~Sakellariadou,
``Is there evidence for topological defects in the BOOMERANG data?,''
Phys.\ Rev.\ D {\bf 65}, 021301 (2002)
[arXiv:astro-ph/0005022].
}

\lref\ru{
A.~Riazuelo, N.~Deruelle and P.~Peter,
``Topological defects and CMB anisotropies: Are the predictions  reliable?,''
Phys.\ Rev.\ D {\bf 61}, 123504 (2000)
[arXiv:astro-ph/9910290].
}

\lref\ruu{
A.~Riazuelo and N.~Deruelle,
Annalen Phys.\  {\bf 9}, 288 (2000)
[arXiv:gr-qc/0005024].
}

\lref\hshs{
G.~T.~Horowitz and A.~R.~Steif,
``Singular String Solutions With Nonsingular Initial Data,''
Phys.\ Lett.\ B {\bf 258}, 91 (1991).
}

\lref\morales{
J.~F.~Morales,
``String theory on Dp-plane waves,''
arXiv:hep-th/0210229.
}


\def\ens{{\it LPTENS,
24 rue Lhomond, 75231 Paris cedex 05, France}}


\Title{\vbox{\baselineskip 10pt  \hbox{LPTENS 02/63}
 \hbox{hep-th/0212217} {\hbox{ }}}} {\vbox{\vskip -30
true pt \centerline{Relativistic String in a Pulse}
\smallskip
\smallskip
\smallskip
   \smallskip\smallskip
\medskip
\vskip4pt }} \vskip -20 true pt \centerline{Constantin  Bachas}
\smallskip\smallskip
\centerline{\ens \foot{Laboratoire mixte du CNRS et de l' Ecole
Normale Sup{\'e}rieure}} 
\bigskip
\bigskip

{\bf Abstract}: I study a relativistic open string coupling
through its endpoints to  a plane wave with  arbitrary temporal profile.
The string's transverse oscillations respond { linearly} to  the
external field. This makes it possible to solve the classical
equations,  and to calculate the quantum-mechanical S-matrix in closed
form. I analyze the dynamics of the string as the characteristic
frequency and duration of the pulse are continuously  varied. I  derive,  
in particular,  the  multipole
expansion in the adiabatic limit of very 
 long wavelengths, and discuss also more violent phenomena 
such as shock waves,  cusps and null brane intersections. 
Apart from their relevance  to the study of time-dependence in 
superstring theory, these results  could have  other applications, such as
the teleportation of gravitational wave bursts by cosmic strings.

\vfil\eject


\newsec{Introduction and Outlook}

Many authors have considered relativistic strings moving in 
the background of a plane-fronted wave  
\refs{\ama\deV\steif\hshs\kk\met\bmn\mett
\atish\gava\kirp\kumar\ske\lms\bgg\Fab\maoz\bain
\fuji\bh\morales\mye\gg\Oku{--}\papart}.  The original motivations
were the study of the high-energy scattering of fundamental strings
\AmatiWQ\ama ,  and of the fate of null spacetime singularities 
\steif\hshs. 
More recently,  interest in pp-wave backgrounds 
has been mainly spurred 
by their conjectured holographic relation to limits of supersymmetric
gauge theories \bmn, and because they provide simple models of time
dependence
in string theory. Finally, 
the interactions  of cosmic strings\foot{For a review of cosmic
strings see \vil\ .}
with electromagnetic and gravitational waves 
may have played an
important role in primordial cosmology, and can be the source of
detectable gravity waves today.

   A plane-fronted gravitational wave is described in Brinkmann
coordinates by the following  metric and curvature tensor~:
\eqn\brinkm{
ds^2 = -2dx^+dx^- + d{\bf x}d{\bf x} +  
H(x^+, {\bf x}) (dx^+)^2\ \ \ \ {\rm and}\ \ \ \ 
R_{i+j+} = \partial_i\partial_j H\ .  
} 
To study a relativistic string in this background, it  is  convenient
to go to the  light-cone gauge, $X^+ = p^+\tau$, 
in which  $H(p^+\tau, {\bf X})$
is  a time-dependent potential for the transverse coordinates. 
The dynamics  simplifies  greatly in the case of  waves with planar
symmetry, also called exact plane waves \grif, for which
$R_{i+j+}$ is constant on the wavefronts.  
In this case   $H = A_{ij}(x^+) X^i X^j$, so that 
the  ${\bf X}(\sigma,\tau)$ are  free worldsheet fields,
with a  mass matrix $A_{ij}$ that depends  only on  time. 
Even this simple problem is, however,  still too hard to solve,  except
for  special time profiles \steif\met\mett\papart\  including 
of course the maximally 
supersymmetric cases   \blau\ where $A_{ij}$ is constant. The difficulty
can be essentially traced to the fact that, 
although the ${\bf X}(\sigma,\tau)$ obey
a linear equation, they have a highly non-local {\it and} non-linear
dependence on the wave profile. 

  In this paper I show that the problem simplifies further
for an open string that couples  to a plane wave
only through  its endpoints.
Such  waves are  electromagnetic rather than
gravitational, or they  may correspond to   geometric vibrations of
D-branes. The striking fact about them is that they enter as 
a  source term in the  field equations for ${\bf X}(\sigma,\tau)$, so
that  transverse string oscillations respond 
{\it linearly}  to the incident wave. This means that one can solve the
classical equations, and 
compute the   exact quantum mechanical S-matrix,
for any  temporal profile of the incident pulse. We will  thus
 study the dynamics of the string
as the  characteristic frequency and duration of the pulse are 
continuously varied. I  will show,  in particular,  how 
the process of string excitation, 
invisible in the long-wavelength  multipole expansion, becomes the
dominant effect for characteristic frequencies of the order of
$(\alpha^\prime p^+)^{-1}$, where $\alpha^\prime$ is the Regge slope
and $p^+$ the conserved light-cone momentum.

   Because of their great  simplicity, these plane waves are privileged
backgrounds  for the study of time dependence in string theory. 
In contrast to other recent works,  where  open strings
have been  attached to D-branes  in a gravitational pp-wave
\atish\gava\kumar\ske\bgg\bain,  
here the worldsheet theory is both massless and free. Furthermore, 
the freedom  to choose the wave profile   makes it possible to 
regularize  both the  ultraviolet and the infrared, so that 
the  first quantized theory has a  well defined,  
albeit  non-trivial S-matrix. This opens the way for    
addressing   a host of other interesting questions~: 
what are the effects of backreaction, and of
open- and closed-string radiation, 
when  string interactions are  turned on?  
 How does the boundary state for an undulating D-brane capture its 
non-trivial time-dependence? What is the  holographic description
of string theory in the  background of a large
collection of undulating branes? On a different register,  
does the linearity of the  coupling to D-brane  plane waves 
impose   constraints  on the  non-abelian extension
of the Dirac-Born-Infeld action?  And finally, what can we learn about
high-energy scattering, and about the fate of cosmological
singularities\foot{My interest in these backgrounds has been
motivated by the analogy between null brane intersections and
orbifold singularities,  advocated in ref. \bh.}
 in string theory? The fact that the worldhseet theory is massless
and free should make some of these questions quite  tractable. 
I hope to return to them  in a forthcoming publication.

   Apart from their relevance to  superstring theory, the
results  of this paper could be also of interest in the context of
large cosmic strings. As I will argue later,  in section 8, the simplified
model studied here  could be relevant  for  a cosmic
string passing near a source of strong gravitational waves. 
The coherent excitation of the string may result in  a teleportation
of the signal, 
assuming  radiation damping can be neglected. This  question deserves,
as I will explain,  further investigation.

The plan of the  paper is as follows~: 

$\bullet$ Section 2 introduces the
basic backgrounds, which are parallel  D-branes with plane Born-Infeld
or vibration waves on their worldvolumes. I explain why these are
exact supersymmetric backgrounds of  superstring theory 
with arbitrary   temporal profiles, and compute their total 
momentum and energy.

$\bullet$  In section 3 we will 
 solve  the classical equations of motion
for  an open  string with both endpoints  on the same undulating D-brane.
I discuss the  point-particle limit, and 
comment on the
similarities between plane-fronted gravitational waves  
and brane waves.  

$\bullet$ Section 4 generalizes the solution  to open strings
with endpoints on distinct D-branes, or in T-dual language to strings
charged with respect to an  electromagnetic pulse. The main
novel feature in this case, is that the wave can impart momentum
to the string,  even in the  limit of extremely long wavelength. 
The reason is that the string  stretches  as the D-branes separate,
or, in T-dual language,   that   charges can   accelerate 
in the  field of a plane electromagnetic pulse. 

$\bullet$ Section 5 presents the quantization of the open
string in the wave background. I calculate the exact S-matrix, paying
particular attention to  zero modes, and derive  the
excitation probabilities and average mass squared of  a string that
was initially in its ground state. The  string emerges 
 in a coherent state,  dictated by the temporal profile of  the pulse.   

$\bullet$ Section 6 is devoted to  the long-wavelength limit of the
classical solution. I derive  a systematic multipole expansion 
in powers of $\alpha^\prime p^+$, or equivalently of  derivatives of the
background fields, and show that it describes a string surfing
on the wave and emerging behind it adiabatically elongated or 
contracted. String excitation cannot be seen at any
finite order of this multipole expansion. 

$\bullet$ In section 7 we will consider  the opposite limit of 
shock waves, and  of waves  with  cusps and kinks. 
We will see how an incident   (primary) short  pulse  
transmits  a (secondary)
clone pulse on  the string, which carries away a fixed fraction
of the incident energy and  momentum.  
We will  also consider other, potentially violent phenomena,
such as resonances and null brane intersections, and comment
on the analogy of the latter 
with null singularities in orbifold models.

$\bullet$ In  section 8 we will ask  under what conditions 
coherent string excitation can be observed in nature.  
I comment on highly energetic cosmic rays in laser beams, and
on large cosmic strings near  strong sources  of gravitational
waves. These latter could be relevant to the experimental search for
gravity waves.

  Let me  finally note  that to keep the  paper short, I 
focus  mainly on the bosonic  degrees of freedom of the string
in what follows. 
Incorporating 
explicitly the fermions does  not present great  difficulties,
and might have obscured our discussion. In the context of
cosmic strings, or of  effective QCD strings, the fermions are 
anyway not required. For fundamental strings, 
on the other hand, supersymmetry will be always   implicitly present. 
It  guarantees that the background configuration is stable,
and that one can try to define a perturbative S-matrix
around it.

\vfil\eject


\newsec{Travelling Waves  on a D-brane}

   Figure 1 shows a D-string with a wave  of arbitrary
temporal profile  travelling at the speed of
light in the (negative) $O1$  direction. Also shown 
in the figure is a reference,  static
and straight D-string. The background spacetime is flat,  
with coordinates 
\eqn\nott{x^\mu = (x^+, x^-, {\bf x}) \ , \ \ \ \
{\rm where}\ \ \ \ x^\pm = {x^0\pm x^1\over \sqrt{2}}\ 
\ \ {\rm and}\ \ \ \ {\bf x}\in \IR^8\ .
} 
 The D-strings 
are described in static gauge by the  embedding 
\eqn\trav{ Y^\pm = \zeta^\pm\ , \ \ \ \ 
{\bf Y} = {\bf Y}(\zeta^+)\ . 
}
Here, and throughout this paper, we use
 $Y^\mu(\zeta^a)$ to denote the spacetime
embedding  of a D-brane, and
$X^\mu(\sigma,\tau)$ for the  embedding  of a fundamental string.
The reference D-string of Figure 1 has ${\bf Y} = 0$, while
for the undulating 
one  ${\bf Y} \equiv  (Y^2, \cdots , Y^9)$ 
is an  arbitrary vector-valued function of $\zeta^+$.

\vskip 0.3cm

\bigskip
\ifig\moverot{A static, straight D-string (left) and a D-string with 
a wave  travelling in the downward direction at the speed of light (right). 
The wiggly lines are open fundamental test strings 
moving  in the  opposite, upward direction.
Since there is neither 
dissipation nor dispersion, the pulse  keeps its  profile and energy
at all times.  After its
passage,   the D-string is  displaced  by ${\bf b}$ in the transverse
space.   
} {\epsfxsize3.3in\epsfbox{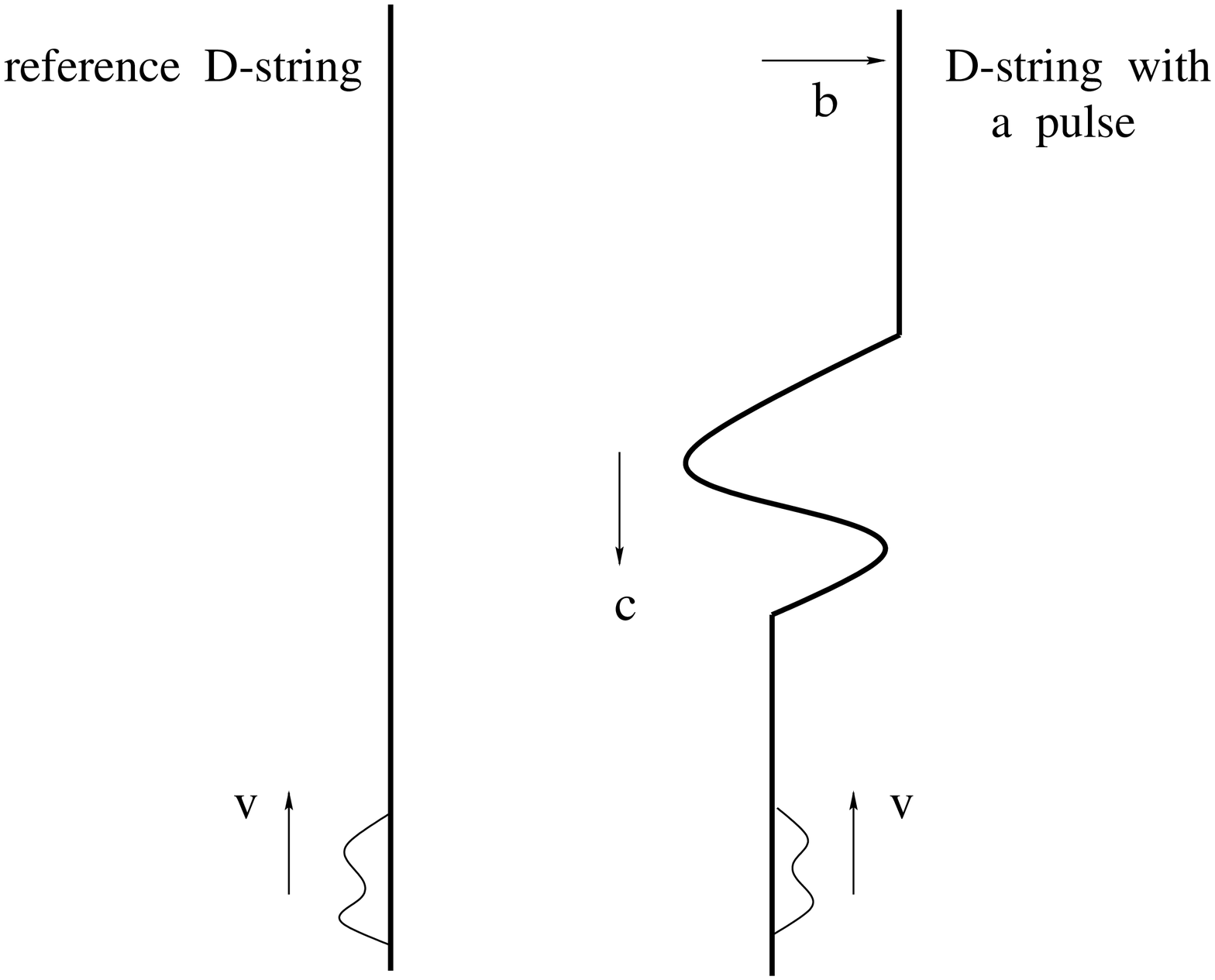}}

\vskip 0.3cm

  The  undulating D-string 
is an exact, consistent background of 
open-string theory for
any choice of the shape  function ${\bf Y}(\zeta^+)$. This may not
seem  obvious at first, because  the effective Dirac-Born-Infeld
equations for a D-brane  receive  
$\alpha^\prime$  corrections,  which generally  depend on the
 extrinsic curvature
\bbg\fot . 
One can, however, argue directly in conformal field theory
that the  boundary operator
\eqn\bnry{ S_{\rm boundary} = \int {d\tau \over 2\pi\alpha^\prime}\;
{\bf Y}(X^+)\cdot \partial_\sigma {\bf X}\ \ , 
}
which describes the coupling of the open string to  the background wave, 
is  marginal for any choice of the profile
function. This is so   because the free worldsheet field
 $X^+(\sigma,\tau)$ has vanishing OPEs   both with itself, 
and with the transverse-coordinate fields  ${\bf X}(\sigma,\tau)$. 
An alternative but equivalent argument is that
all potential contributions  
to the spacetime equations
for ${\bf Y}(\zeta^a)$ are zero,  
because  terms having  a lower `minus' index vanish, 
while  those without a minus index cannot be contracted.

   Actually  the  above configuration is not only consistent,
but it is also  1/4 supersymmetric. If  $\epsilon$ and
$\bar\epsilon$ are the two Weyl-Majorana supersymmetry parameters of
type IIB theory, then the  unbroken supersymmetries obey
\eqn\susy{ \Pi \epsilon = \bar \epsilon \ \  
\ \ \ {\rm and} \ \  \ \  \gamma^+\epsilon = 0\ , 
}
where $\Pi$ is the product of $\gamma$ matrices
along the directions of the reference brane  (up to a phase \rev). 
Supersymmetric strings with travelling waves have been 
discussed widely in the literature, both in the supergravity 
approximation \har\ , 
and more recently as consistent configurations of D-branes 
\mat\cho\lun\hya\ . They  play
an important role in the study of string dualities, and in the statistical
description  of near-extremal black holes. 

  The wave illustrated in Figure 1 dies out asymptotically,
meaning that  ${\bf Y}$ goes to constant values in the distant future and 
in the distant past. This guarantees that the pulse has finite energy, and
that we  can define asymptotic states for the scattering problem. 
By an appropriate choice of  the origin of coordinates we can set
\eqn\asympt{ {\bf Y}(-\infty) = 0\ \ \ \ {\rm and}
\ \ \ \  {\bf Y}(+\infty) = {\bf b}\ .
}
We will relax this condition,  and consider infinite wave trains towards
the end of this paper.  Until further notice  \asympt\ will be,  however, 
implicitly  assumed.

By the standard rules of T-duality  (see for instance \rev)
one  can map the undulating D-string to an equivalent higher-dimensional
D-brane, with  a plane electromagnetic wave on its worldvolume, 
\eqn\emwave{ {Y}^j(\zeta^+) \; \equiv \; 
2\pi\alpha^\prime {A}^j (\zeta^+)\ .} 
This  is a homogeneous,  generally  non-monochromatic, 
electromagnetic wave. In what follows, 
I will mostly use the geometric language of undulating
D-branes,  which is easier to visualize. All the results will, 
however,  apply with little change to plane
electromagnetic waves interacting
with the endpoints of an open string.

    To  conclude this section,  let us calculate the energy and momentum
carried by a  pulse. To leading order in the $\alpha^\prime$ expansion this
is given by the variation of the Dirac-Nambu-Gotto action for the D-string,
\eqn\emom{T^{\mu\nu}(x) = -{2 T_D } \int d^2\zeta\; 
{\partial \sqrt{-\hat g (\zeta)} \over \partial  G_{\mu\nu}(x)}\ \ ,
} 
where $\hat g$ is the determinant of the induced metric,  $T_D$ the tension
of the D-string, and $G_{\mu\nu}$ the spacetime metric. 
Using the identity 
\eqn\id{  {\partial\; \hat g (\zeta)  \over \partial  G_{\mu\nu}(x)}\;
= \; \hat g (\zeta)\; {\hat g}^{ab}\; \partial_aY^\mu\partial_bY^\nu\;  
\delta^{(10)}\left(x^\rho -Y^\rho(\zeta)\right)\; 
\ \ ,
}
and the special form of the induced metric in our problem, 
\eqn\indmetr{ \left( \matrix{ \hat g_{++} & \hat g_{+-}\cr & \cr  
\hat g_{-+} & \hat g_{--} \cr} \right)\;  = 
\; \left( \matrix{ \partial_+{\bf Y}\cdot \partial_+{\bf Y}  & -1 \cr
 &  \cr -1  & 0 \cr} \right)
\ , 
}
one finds the following non-vanishing components of the energy-momentum tensor
\eqn\emo{
\matrix{ T^{+-}\cr  T^{--}\cr T^{-j}\cr}
\Biggr\}  \; = \; T_D \;
\delta^{(8)}({\bf x}- {\bf Y}(\zeta))\;\times\;  \Biggl\{
\matrix{ 1\ \ \  \cr \partial_+ {\bf Y}\cdot 
\partial_+ {\bf Y} \cr \partial_+ {Y}^j \ \cr}
\ \ \  . }
Note that the right-hand side is a function  of $x^+ = \zeta^+$. 
Subtracting the energy-momentum tensor of the static reference D-string,
and integrating over $\IR^9$,  gives the total energy and momentum of 
the pulse~: 
\eqn\enmo{ P^+_{\rm pulse} = 0\ \ \ {\rm and}\ \ \ \ 
P^-_{\rm pulse} = {T_D } \int {dx^1\over \sqrt{2}} \;
\partial_+ {\bf Y}\cdot \partial_+ {\bf Y}
\ .
}
Finiteness of the energy requires that ${\bf Y}$ grows at most
like $\vert x^+\vert^{1/2}$ at infinity, a weaker condition  
than the assumption \asympt\ that we already made.

The semiclassical configuration \trav\ is  a coherent
superposition  of  open strings, moving in the
downward direction at the speed of light. In terms of the 
 Fourier components of  ${\bf Y}$, 
\eqn\four{ {\bf Y}(x^+) \equiv  \int_{-\infty}^\infty
 {dp^-\over  p^-}\;   
{\bf y}(p^-)\; e^{ip^- x^+}\ ,   
}
the total energy-momentum of the pulse  reads  
\eqn\fou{ P^-_{\rm pulse} = 4 \pi T_D \int_0^\infty dp^- \; 
 \vert {\bf y}(p^-)\vert^2\ \ \ . 
}
The  average density of open strings with momentum $p^-$ is therefore
\eqn\densi{ p^- < n(p^-)>\; =\; 4\pi T_D\;  \vert {\bf y}(p^-)\vert^2\ . 
}
Since $T_D\sim 1/g_s$,  this  density 
is very large at weak string coupling. Fluctuations
in the  density are  subleading, $ \delta n \sim o(\sqrt{n}) $, 
consistently  with the semiclassical treatment  of the
pulse.  Note that if ${\bf Y}$ dies out more slowly  than $\sim 1/x^+$ at
infinity, the average number density of open strings 
diverges in the infrared.


\newsec{Neutral Test Particle  and  String}

 We turn next  to the analysis of open strings moving
in the background of
such  travelling waves. There are two types of open strings~: 
(a) those illustrated in Figure 1, which have 
both endpoints on the same D-brane, and (b) those with endpoints
on two distinct D-branes (see Figure 2). 
In the T-dual language, the former correspond to  
 neutral strings, while the latter are charged
with respect to the electromagnetic wave. 
I  will consider  the  neutral case first, and  extend the analysis
to charged strings  in the following section. I will also first 
solve  the classical problem, and then   proceed  to quantize 
the  string  in section 5. 
The study  of the  second-quantized theory, and in particular of 
the backreaction of the string on the travelling wave,  
is  postponed to a future publication.

 Before analyzing relativistic strings, it is instructive to first consider a
point particle confined to move on the D-string of Figure 1. The net effect 
of the pulse on the particle
is   a time delay,   and  
a  displacement   ${\bf b}$ in the transverse space. 
Nothing else can  happen  to the particle, since (a) its momentum 
$p^+$ is conserved, (b) there are no
internal degrees of freedom to excite, 
and (c) the mass-shell condition  $2p^+ p^- = m^2$ implies that $p^-$
is fixed in the asymptotic past and future. To  calculate 
the time delay
of the particle, let us define 
 flat coordinates on the D-string,
\eqn\flatc{\zeta^+= x^+\ \ \ \  {\rm and}\ \ \ \ 
\zeta^- = x^- - {1\over 2}\; \int_{-\infty}^{\; x^+}\; 
 (\partial_+{\bf Y})^2\ \ . 
}  
In these coordinates $\hat g_{\alpha\beta}= \eta_{\alpha\beta}$ 
and  the particle's  worldline is a straight line~:
$\zeta^\pm = p^\pm \tau$. Thus, in terms of
the original  spacetime coordinates we have~:
\eqn\lin{ X^+(\tau)  = p^+\tau\ \ \ {\rm and}\ \ \ \ 
X^-(\tau) = p^-\tau + {1\over 2}\; \int_{-\infty}^{\; p^+\tau}\; 
 (\partial_+{\bf Y})^2\ \ . 
} 
Comparing with  the worldline on a straight reference
D-string, for which $X^\pm = p^\pm \tau$, we see that
the passage of the wave leads to a net total shift of $X^-$ by an amount
\eqn\delay{\delta X^- = {1\over 2}\; \int_{-\infty}^{+\infty}\; 
 (\partial_+{\bf Y})^2 \; =\; {P^-_{\rm pulse}\over 2T_D}\ . 
}
The second equality in this expression follows from eq. \enmo\ . 
The  positive-definite  shift of  $X^-$ implies that the particle on the
undulating string reaches any given  point far behind the pulse later than
if it were moving on the reference brane. 

   The above results are strongly reminiscent of the physics of gravitational
waves. Particles crossing,  for example, 
 a gravitational shock wave \as\thooft\ 
experience a shift 
$\delta X^- \propto G_N P^-$, reminiscent of  the
delay \delay\ . 
The passage of a gravitational wave burst  can cause,  furthermore, 
a  net transverse displacement,  like
 ${\bf b}$, 
an effect known as  the `memory' of the wave \zel\chri. 
These similarities
should not  come as  a  surprise, since our  point particle 
couples to the induced metric \indmetr\ , which has precisely  the
Brinkmann form of a plane-fronted wave.  
The reader may here object that in 
a real gravitational wave $g_{++}$ must depend on  transverse 
coordinates, since otherwise $R_{i+j+}=0$ 
and  the spacetime is everywhere   flat.  
The  transverse space is, however,  effectively
 `deconstructed' in our
case  by juxtaposing several D-strings, each  with a different  wave 
profile ${\bf Y}(x^+)$. The time delay \delay\ ,  in particular,  
arises  because we compare 
 motion on the  undulating D-string,   to motion
on a static  reference D-brane.


 Consider next  a  relativistic open string in the background
of the travelling wave. The 
embedding coordinates of the open string obey the usual free-wave equation 
\eqn\freew{ ( \partial_\tau^2  - \partial_\sigma^2)
  X^\mu(\sigma,\tau) = 0\ , 
}
and Virasoro conditions
\eqn\vira{ \partial_+X^\mu \; \partial_+X_\mu\; =\; 
\partial_-X^\mu \; \partial_-X_\mu\; =\; 0 \ ,
}
where $\partial_\pm = {1\over 2}(\partial_\tau\pm \partial_\sigma)$. 
Since the center-of-mass momentum $p^+$ is conserved, we can go 
to the lightcone gauge in which 
\eqn\lightc{ X^+ = 2\alpha^\prime p^+\tau \ \ \ {\rm and}\ \ \ \ 
2\alpha^\prime p^+ \partial_\pm X^- = 
 \partial_\pm{\bf X}\cdot \partial_\pm{\bf X}\ \ .
}
We  set from now on $2\alpha^\prime =1$. 
As usual, one can  solve the Virasoro  conditions 
for $X^-(\sigma,\tau)$ in terms of $p^+$ 
and of the transverse coordinates ${\bf X}(\sigma,\tau)$. 
The  latter  must obey  the boundary conditions
\eqn\bcond{ {\bf X}(\sigma,\tau) = {\bf Y}( p^+\tau) \ \ 
\ {\rm at}\ \ \sigma = 0\ \ {\rm or}\ \  \pi\ \ , 
}
since  both  string endpoints 
lie on the worldline of the undulating D-string. 
Equations \freew\ and \bcond\ define a linear problem, in which
the wave profile ${\bf Y}(p^+\tau)$  enters as a { source}. 
This should
be contrasted to  other electromagnetic backgrounds \refs{\elo\eltwo
{--}\elt}\  
which,  even when they leave the string equations linear, 
modify at the  least their homogeneous part.

The  general solution to the above  linear problem 
can be written in two  different ways as follows~:
\eqn\inout{ {\bf X}\; =\; {\bf X}_{\rm in} + \delta {\bf X}_{\rm in}\; =\;
{\bf X}_{\rm out} + \delta {\bf X}_{\rm out}\ , 
}
where ${\bf X}_{\rm in\; (out)}$ 
are  general solutions  to the  
homogeneous problem long before (or after) the passage of the pulse, 
and $\delta {\bf X}_{\rm in\; (out)}$  
are special  solutions (up to a constant) 
that vanish at $\tau\to -\infty\; (+\infty)$. 
Since  ${\bf X}_{\rm in}$ and ${\bf X}_{\rm out}$
obey  standard Dirichlet conditions
 at both string endpoints, 
their mode expansions  are of the usual form~: 
\eqn\gesol{ {\bf X}_{\rm in}(\sigma,\tau)\; = \; 
\sum_{n\not=  0} 
\; {1\over n}\; {\bf a}_n^{\rm (in)} \; e^{-in\tau}\; {\rm sin}(n\sigma)\ ,
}
and 
\eqn\gesolu{ {\bf X}_{\rm out}(\sigma,\tau)\; = \; {\bf b}\; +
\; 
\sum_{n\not=  0} 
\; {1\over n}\; {\bf a}_n^{\rm (out)} \; e^{-in\tau}\; {\rm sin}(n\sigma)\ . 
}
Our main  new result is  the following exact  expression for the special
solutions~:
\eqn\spec{ \delta {\bf X}_{\rm in}(\sigma ,\tau)\; = \;
 {\bf F}_{\rm in}(\tau -\sigma , \; p^+)
\; +\; {\bf F}_{\rm in} ( \tau +\sigma -\pi ,\;  p^+) \ \ , 
}
and similarly for  `in' replaced by `out',
where the functions ${\bf F}_{\rm in\; (out)}$ are 
defined  by the following alternating sums~: 
\eqn\speclc{ {\bf F}_{\rm in}(a, \lambda)\; = \;\sum_{N=0}^\infty\; (-)^N \;
 {\bf Y}(\lambda a  - \lambda N\pi) \ ,  
}
and 
\eqn\specia{ {\bf F}_{\rm out}(a, \lambda)\; = \;- \sum_{N=1}^\infty\; (-)^N \;
\left[ {\bf Y}(\lambda a  + \lambda N\pi) - {\bf b} \right] \ . 
}
To check that the boundary conditions \bcond\ are 
indeed satisfied, the reader should
note that the function defined in \speclc\   
 obeys  the relation ${\bf F}_{\rm in}(\tau, p^+) + 
{\bf F}_{\rm in}(\tau-\pi, p^+)
= {\bf Y}(p^+\tau) $, and  that
a similar relation holds for the function  ${\bf F}_{\rm out}$.   
 Furthermore, for  $p^+$ strictly  positive, 
$\delta {\bf X}_{\rm in}$
vanishes in the asymptotic  past provided
that  ${\bf Y}$ goes to zero fast enough,
and the  same is true 
for $\delta {\bf X}_{\rm out}$  in  the asymptotic future.

Equations \inout--\speclc\ are the complete solution to  the 
initial-value problem, whenever   the alternating sum 
${\bf F}_{\rm in}(a,\lambda)$ can be defined.  Note that in the 
$p^+\to 0$ limit,  the  solution reduces to  a point-like string 
 riding   the
wave  at constant values of $x^+$ and of ${\bf X} = {\bf Y}(x^+)$. 
When  $p^+\not=0$ the string collides inevitably 
with the travelling wave, and  will emerge   in general
with  different   momentum $p^-$ and mass $M$. We will analyze 
this process in detail, 
for various  wave profiles and in different physical contexts, 
in the following sections. First, however, I  will 
generalize the above  solution to the case of an open string
stretching  between two different D-branes.


\newsec{Charged  Particle and String}  

    This  more general  situation is  illustrated in Figure 2. The 
oriented open string
has its left endpoint on one D-string, and its right endpoint on another.
The travelling waves on these  two D-strings are  a priori  different,
and can be  decomposed  into a synchronized  and a relative motion~:
\eqn\twods{
  {\bf Y}_{\rm left} = {\bf Y}(x^+) +  {\bf Z}(x^+)\ \ \ \ {\rm and}
 \ \ \ \ {\bf Y}_{\rm right} = {\bf Y}(x^+) -  {\bf Z}(x^+)\ . 
}
This configuration is still a consistent, 1/4 supersymmetric background
of open-string field theory. 
 The boundary conditions in light-cone gauge are~:
\eqn\bcondd{ {\bf X}(0,\tau) =
 {\bf Y}_{\rm left}( p^+\tau)\ \  \ {\rm and}
 \ \ \ \ {\bf X}(\pi,\tau)= {\bf Y}_{\rm right}( p^+\tau)\ . 
}
Clearly, in the absence of the  relative wave the problem
reduces to the one of the previous section. Our task here is to generalize  the
solution to the case of non-trivial  ${\bf Z}(x^+)$.

\bigskip 
\ifig\probl{An open string stretching between two D-branes  
with  different plane-waves on their worldvolumes. 
The passage of the wave train 
can change the distance between the two D-branes, and hence also the 
minimal mass of the string. In the 
 T-dual configuration this corresponds to a  charged string
picking up transverse momentum 
 in the electric field of the  wave. 
} {\epsfxsize2.2in\epsfbox{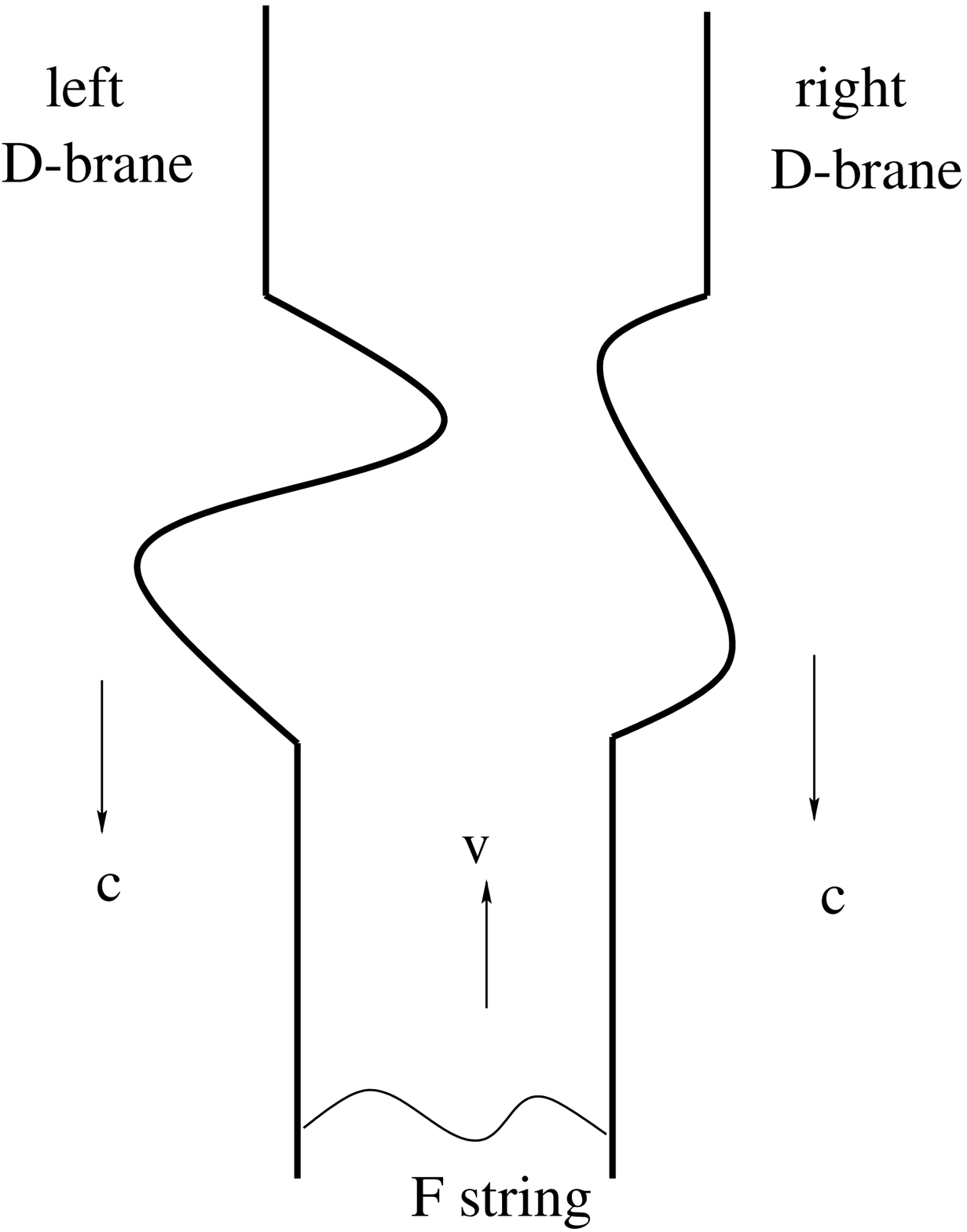}}

The general  solution can be again written 
in the form \inout\ , where ${\bf X}_{\rm in\; (out)}$ are the 
asymptotic fields that obey  conventional Dirichlet boundary conditions, 
and $\delta {\bf X}_{\rm in\; (out)}$  vanish  in the
distant past (or  future) as before. To be explicit, let us assume
that ${\bf Y}$ has the asymptotic limits \asympt\ , while   
\eqn\asymptt{ {\bf Z}(-\infty) = \pi {\bf w}_{\rm in} \ \ \ \ {\rm and}
\ \ \ \  {\bf Z}(+\infty) = \pi {\bf w}_{\rm out}\ ,
}
so that the passage of the pulse
changes the inter-brane distance  from $2\pi{\bf w}_{\rm in}$
to $2\pi{\bf w}_{\rm out}$.  
Then the asymptotic `in' and `out' fields read~:
\eqn\gesoll{ {\bf X}_{\rm in}(\sigma,\tau)\; =  
\; {\bf  w}_{\rm in}\; (\pi   - 2{\sigma})\; +\;
\sum_{n\not=  0} 
\; {1\over n}\; {\bf a}_n^{\rm (in)} \; e^{-in\tau}\; {\rm sin}(n\sigma)\ ,
}
and  
\eqn\gesolll{ {\bf X}_{\rm out}(\sigma,\tau)\; =\; {\bf b}\; +  
\; {\bf  w}_{\rm out}\; (\pi   - 2{\sigma})\; +\;
\sum_{n\not=  0} 
\; {1\over n}\; {\bf a}_n^{\rm (out)} \; e^{-in\tau}\; {\rm sin}(n\sigma)\ . 
}
Furthermore, since the problem is linear in the external fields,  
the special solutions are those given in
  the previous section,  plus an extra 
contribution that takes care of the relative 
motion of the branes ${\bf Z}(x^+)$.
 Explicitly~:
\eqn\speccc{ \matrix { \delta {\bf X}_{\rm in}(\sigma ,\tau)\; = \;
 {\bf F}_{\rm in}(\tau -\sigma , \;  p^+) \; +\; 
{\bf F}_{\rm in} ( \tau +\sigma -\pi ,\;  p^+) 
  \cr  \cr  
\ \  \ \  \ \ \ \  \ \  \ \ \  \ \  \ \ \ \  \; +\;    
{\bf G}_{\rm in}(\tau -\sigma , \;  p^+)
\; -\; {\bf G}_{\rm in} ( \tau +\sigma -\pi ,\;  p^+)\ \ ,   \cr} 
} 
\vskip 0.2cm\noindent
where ${\bf F}_{\rm in}$ is defined as in \speclc\ , and  
\eqn\specl{ {\bf G}_{\rm in}(a, \lambda)\; = \;\sum_{N=0}^\infty\; 
\left[ {\bf Z}(\lambda a  -
 \lambda N\pi) -  \pi {\bf w}_{\rm in} \right] \ .   
}
Similarly 
\eqn\speclc{ \matrix { \delta {\bf X}_{\rm out}(\sigma ,\tau)\; = \;
 {\bf F}_{\rm out}(\tau -\sigma , \;  p^+) \; +\; 
{\bf F}_{\rm out} ( \tau +\sigma -\pi ,\;  p^+) 
  \cr  \cr  
\ \  \ \  \ \ \ \  \ \  \ \ \  \ \  \ \ \ \  \; +\;    
{\bf G}_{\rm out}(\tau -\sigma , \;  p^+)
\; -\; {\bf G}_{\rm out} ( \tau +\sigma -\pi ,\;  p^+)\ \ ,   \cr} 
} \vskip 0.1cm  
\noindent where  ${\bf F}_{\rm out}$ is given by equation 
\specia\ , and 
\eqn\specia{ {\bf G}_{\rm out}(a, \lambda)\; = \;- \sum_{N=1}^\infty\; 
\left[ {\bf Z}(\lambda a 
 + \lambda N\pi) -  \pi {\bf w}_{\rm out}\right] \ . 
}
The  new boundary conditions \bcondd\ are 
 satisfied thanks to the  identity~:
 ${\bf G}_{\rm in}(\tau, p^+)
- {\bf G}_{\rm in}(\tau-\pi, p^+) = {\bf Z}(p^+\tau)$,  which  the reader 
will have no trouble to verify. Clearly  the  solution \speccc\ 
  reduces   to the one   of the previous
section when ${\bf Z}(x^+)$,  and 
hence also the sums ${\bf G}_{\rm in}$ and ${\bf G}_{\rm out}$, 
 are set to zero.

     It is instructive to dualize the above solution, so as to  
describe  an open string coupled  to Born-Infeld
electromagnetic  waves. The T-dual gauge fields are 
 \eqn\doublt{ {\bf A}\pm {\bf B}
 \;  \equiv  \; (2\pi\alpha^\prime)^{-1}({\bf Y}\pm {\bf Z}) 
 \ \ .
}   
The open string is charged  with respect to the  field
$B$, and couples as a dipole to the  field $A$.  
The   classical string trajectory is readily
obtained by changing the sign of the right-moving part of the
transverse coordinates,
 $\partial_{\pm} {\bf X} \to \mp \partial_{\pm} {\bf X}$ . This is
 the action of a T-duality transformation on the worldsheet \rev.
In the case at hand we find ${\bf X}^\prime = {\bf X}_{\rm in}^\prime
 +\delta {\bf X}_{\rm in}^\prime$, with
\eqn\tdone{ {\bf X}_{\rm in}^\prime (\sigma,\tau)\; =  
\; {\bf  p}_{\rm in}\;\tau \; -\;
\sum_{n\not=  0} 
\; {i \over n}\; {\bf a}_n^{\rm (in)} \; e^{-in\tau}\; {\rm cos}(n\sigma)\ 
}
and 
\eqn\tdtwo{\matrix { \delta {\bf X}_{\rm in}^\prime(\sigma ,\tau)\; = \;
 {\bf F}_{\rm in}^\prime(\tau -\sigma , \;  p^+) \; - \; 
{\bf F}_{\rm in}^\prime ( \tau +\sigma -\pi ,\;  p^+) 
  \cr  \cr  
\ \  \ \  \ \ \ \  \ \  \ \ \  \ \  \ \ \ \  \; +\;    
{\bf G}_{\rm in}^\prime(\tau -\sigma , \;  p^+)
\; + \; {\bf G}_{\rm in}^\prime ( \tau +\sigma -\pi ,\;  p^+)\ \ ,    \cr} 
}
where  
\eqn\spedu{ {\bf F}_{\rm in}^\prime(a, \lambda)\; = \;
\pi\;\sum_{N=0}^\infty\; (-)^N \;
  {\bf A}(\lambda a  - \lambda N\pi) \  
}
and 
\eqn\specdu{ {\bf G}_{\rm in}^\prime(a, \lambda)\; = \;\pi
 \;\sum_{N=0}^\infty\; 
 {\bf B}(\lambda a  -
 \lambda N\pi)   \ .   
}
We have here used  a gauge
transformation to  set  ${\bf A}(-\infty)= {\bf B}(-\infty)=0$.
Notice that the asymptotic  brane separation $2\pi {\bf w}_{\rm in}$ 
has been  mapped here to  transverse momentum in the initial state.    
Equations \tdone\ -- \specdu\ are the complete 
solution to  the problem of an open string
coupling to two parallel 
plane  electromagnetic waves.

   Before discussing 
further  the above  solution, it will be again useful to comment on
its  point-particle limit. The novelty  with respect to our 
discussion in  the previous section, is that a  charged point particle 
can now  accelerate in the electric field of the wave ${\bf B}(x^+)$,
and emerge with a modified momentum
\eqn\modmom{{\bf p}_{\rm out} -  {\bf p}_{\rm in}\;
 =\; - q\; {\bf B}(+\infty)\ . 
}
This follows by integrating the  equation of motion
\eqn\stpoint{m {d^2X^\mu\over d\tau^2} =
 q F^{\mu\nu}{dX_\nu\over d\tau} \ ,    
}
where  $F_{+j}= \partial_+ B_j$,  
and $q$ is the charge of the particle that I  assumed   minimally coupled.  
Since $p^+$ is always conserved,  and  
$p^- = ({\bf p}^2+m^2)/2p^+$ in the asymptotic regions, 
it follows that the momentum $p^-$ 
will also generally change  in the process.
This  possibility of net  momentum transfer even 
in the point-particle limit
is the main difference between a charged and  a neutral string. 
Notice that for  a neutral particle,  the invariant
momentum transfer $t = (p_{\rm out} -p_{\rm in})^2$ is zero, because both
$p^+$ and the transverse ${\bf p}$ are  conserved.

    In the original T-dual picture of Figure 2,  momentum transfer
corresponds to the elongation  (or contraction)  of the open
string, due to  the net relative displacement of  the D-branes. 
 This duality has been  invoked previously in  \bpio\ 
in order to
give a `soap-bubble' interpretation of the high-energy 
behaviour of the Veneziano amplitude.
 Note also that the above analysis  can be
extended easily to  `mixed'  situations, such as the   Dp/D(p+4) system  
carrying   both vibration and 
electromagnetic waves.
 Rather than guessing the special solutions,
as we have done here, it is however preferable to  construct them
directly using the  Green functions  on the worldsheet. This
brings us naturally to the issue of first quantization, 
to which we  will next turn our attention.


\newsec{First Quantization}

   The quantum mechanical evolution of the string is described  by
a unitary operator  acting on the asymptotic incoming Fock space.
This Fock space is constructed as usual, by interpreting the oscillation 
amplitudes ${\bf a}_n^{(in)}$ of the incoming fields
 \gesoll\ as  creation
or  annihilation operators. In the 
 interaction picture,  the evolution operator 
 follows directly from the expression \bnry\
 for  the marginal deformation,   
\eqn\evolt{ U(\tau)\;  =\;  T\;  {\rm exp}\left( i\int_{-\infty}^{\; \tau} {
d\rho\over \pi}\;  {\cal H}_{\rm I}(\rho)\;    \right)\ ,
}
\noindent where $T$ denotes time ordering, and 
the interaction Hamiltonian is
\eqn\inth{
{\cal H}_{\rm I}(\rho)  = {\bf \tilde Y}_{\rm left}( p^+\rho)\cdot 
\partial_\sigma {\bf X}_{\rm in}(0, \rho) -
{\bf \tilde Y}_{\rm right}( p^+\rho)\cdot \partial_\sigma
 {\bf X}_{\rm in}(\pi, \rho)\ . }
Here the tildes signify  that  the past asymptotic values
of  the displacement vectors, i.e.  
$\pm \pi {\bf w}_{\rm in}$ for ${\bf Y}_{\rm left}$ and ${\bf  Y}_{\rm right}$,
are  subtracted away. Using the definitions \twods, and 
the expansion of the incoming field  \gesoll, one finds
\eqn\inth{
{1 \over 2}\; {{\cal H}_{\rm I}(\rho)}  =  
 \sum_{n\ {\rm odd}}
e^{-in\rho}\; {\bf a}_n^{\rm (in)} \cdot {\bf Y}( p^+\rho)\;
+\; \sum_{n\ {\rm even}} 
 e^{-in\rho}\; {\bf a}_n^{\rm (in)}\cdot
 \left( {\bf Z}( p^+\rho) -
 \pi {\bf w}_{\rm in} \right)\;  \ ,
}
where we have defined the zero mode $
{\bf a}_0^{\rm (in)}  \equiv -2 {\bf w}_{\rm in}$, and the dots
denote the vector inner product. 
The worldsheet S-matrix, that maps the incoming  states of the 
open string to outgoing single-string states, can be  obtained as the limit
of the evolution operator, 
\eqn\smatr{ S\; =\; {\rm lim}_{\tau\to\infty}\; U(\tau)\ .
}

  Since the interaction is linear in the incoming fields, whose
commutators are just  c-numbers,  we can remove the time ordering 
of  $U(\tau)$ at the expense of introducing an overall
{\it real} phase  (see for instance \iz )~: 
\eqn\izone{  U(\tau)\;  =\;  {\rm exp}\;(i\delta_1)\;\times\; 
 {\rm exp}\left( i\int_{-\infty}^{\; \tau} {
d\rho\over \pi}\;  {\cal H}_{\rm I}(\rho)\;    \right)\ ,
}
where
\eqn\iztwo{\delta_1 = {i\over 2}\int_{\infty}^\tau\int_{\infty}^\tau
{d\rho\; d\rho^\prime \over \pi^2}\;
 [\;{\cal H}_{\rm I}(\rho),\; {\cal H}_{\rm I}(\rho^\prime)\;]\;
 \theta(\rho -\rho^\prime)\;\ .
}
Note that ${\cal H}_{\rm I}(\rho)$ is a Hermitean operator, so $\delta_1$ is
indeed a real phase. It is a function of the momentum $p^+$, and of
the charge sector of the open string, but does not depend otherwise on
the incoming state. For a neutral string in the adiabatic limit
(see next section) this  phase shift should be proportional
to  the time delay \delay\ . More generally, it can be evaluated  
explicitly, but  this  will not concern us any further here.

 Let us instead consider   the string-coordinate operators, which 
can be expressed  in
 the Heisenberg representation  as follows~:
\eqn\heis{ {\bf X}(\tau , \sigma)\; \equiv  \;{\bf X}_{\rm in}(\tau , \sigma)+
\delta {\bf X}_{\rm in}(\tau , \sigma) = 
 U(\tau)^{-1}\; {\bf X}_{\rm in} (\tau , \sigma )\; U(\tau)\;\ .  
} 
Using the equation \izone\   and the identity 
\eqn\ident{ e^{-A}\; B\; e^A\; =\; B\; +\; [A,B]\ ,
}
valid whenever  the commutator $[A,B]$ is a c-number, one finds 
\eqn\deltax{ \delta {\bf X}_{\rm in}(\tau , \sigma)\; =\; 
-i \int_{-\infty}^{\; \tau} {
d\rho\over \pi}\;
[\; {\cal H}(\rho) ,\;  {\bf X}_{\rm in}(\tau , \sigma)\; ]\ \ .
}
Plugging in the expression \inth\ for the interaction Hamiltonian,
and using the  canonical commutation relations  
\eqn\comm{ [ a_m^i, a_n^j]\; =\; m\;\delta_{m+n}\;\delta^{ij}\ , 
}
as well as the identities
\eqn\identt{ 
\sum_{n\ {\rm even}} e^{in \alpha } = \pi \sum_{m= -\infty}^{\infty}
\delta(\alpha -m\pi)\ \  ,} 
and  \ \ \ 
\eqn\identtt{ 
\sum_{n\ {\rm odd}} e^{in \alpha } = \pi \sum_{m= -\infty}^{\infty}
(-)^m \; \delta(\alpha -m\pi)\ \  ,
}
gives precisely   
\foot{The contribution of the zero mode in this calculation is subtle. It
arises because in the  expansion  \gesoll\ for the quantum field, one
must include  the dual position operator that has a canonical commutation 
relation with   ${\bf a}_0$.} 
the expression \speccc\  for $\delta {\bf X}_{\rm in}$. 
In fact, the above argument is  an alternative, direct
 derivation of this
expression, which can be extended easily to the general
`mixed' situations mentionned in section 4. 
It is of course  not surprising that 
the semiclassical approximation is exact, since the interaction Hamiltonian
is linear. 


We concentrate now on the  $S$ matrix, which describes the outgoing  states
of the open string long after the passage of the wave.
Combining equation \inth\ with our definition \four\  of
the Fourier transforms  we find~:  
\eqn\transone{ i \int_{-\infty}^{\; \infty}{d\rho \over \pi}\; {\cal H}_{\rm I}
\; =\;  i {h}_0\; +\;
\sum_{n \; {\rm odd}}\; {4i\over n}\; {\bf a}_n^{\rm (in)} \cdot 
{\bf y}(n/p^+) \ + \ 
\sum_{n  \; {\rm even}\not= 0}\; {4i \over n}\; {\bf a}_n^{\rm (in)} \cdot 
{\bf z}(n/p^+)\ \ , 
}
where ${\bf y}$ and ${\bf z}$ are the Fourier transforms, respectively,
of ${\bf Y}$ and ${\bf Z}$, and $h_0$ is 
the contribution of the zero modes, 
\eqn\auxx{
h_0 =  {2 } \int_{-\infty}^\infty
 {d\rho\over\pi}  \; {\bf a}_0^{\rm (in)}\cdot
 ({\bf Z}(p^+\rho) -\pi {\bf w}_{\rm in})\ \ .   
}
Notice that ${\bf a}_0^{\rm (in)}$
 is here  an  operator, while ${\bf w}_{\rm in}$ is the
asymptotic value of the wave. The integral \auxx\  diverges  whenever
${\bf w}_{\rm out} \not= {\bf w}_{\rm in}$, i.e. when the 
string is elongated (or contracted) by the wave. 
In this case $h_0 = 2 ({\bf w}_{\rm out}-
{\bf w}_{\rm in})\cdot {\bf \hat x}_0$ , where
${\bf \hat x}_0$ is the pseudo-position operator 
that is canonically conjugate 
to ${\bf a}_0^{\rm (in)}$. This  statement can be  understood 
easily if one considers 
 a charged particle in an electromagnetic pulse, for which
the S-matrix must obey  the relation $S^{-1}{\bf \hat p}_{\rm in} S =
 {\bf p}_{\rm out} - {\bf p}_{\rm in}$.

Let us consider finally
 the oscillator contributions to the S-matrix.
Using the well-known   identity, valid  for a c-number commutator, 
\eqn\newid{ e^{A+B}\; =\; e^A\; e^B\; e^{[A,B]/2}\ , 
}
 we can  put the S-matrix in normal-ordered form~:  
\eqn\smatrix{ S\; =\;   {\rm exp}\;(i\delta_1 -\delta_2)\;\times\; 
:  {\rm exp}\left( i\int_{-\infty}^{\; \infty} {
d\rho\over \pi}\;  {\cal H}_{\rm I}(\rho)\; \right) : \ \ \ , 
}
where the imaginary phase
$\delta_2$ can be extracted easily from the expression 
\transone\ , 
\eqn\deltatwo{
\delta_2 \; =\; 
\sum_{n=1,3,  ..} \; {8\over n}\; 
\vert\;  {\bf y}(n/p^+)
\;\vert^{\; 2}
\;+\; 
 \sum_{n=2,4,..}\; {8\over n}\;  
\vert \; 
 {\bf  z}(n/p^+)\; \vert^{\; 2} 
\; \ \ .  
}
Equations  \transone -- \deltatwo\ are the main result of this
section. They give the exact 
S-matrix of the  open quantum  string in the background of 
an arbitrary plane wave.

 A  string  initially in its ground state  will emerge
in the coherent superposition
 $S \vert\;  p^+ >_{\rm  in}$ after crossing the
pulse.  The imaginary phase $\delta_2$ determines the probability 
for the  string  not to be excited by the wave,     
\eqn\pzer{ \vert \;_{\rm in} \kern-.29em
< p^+\vert\; S\;\vert\; p^+ >_{\rm  in}
\vert^{\; 2} \;
=\; e^{-2\delta_2}\ .   
}
Notice that the synchronized wave,  ${\bf Y}(x^+)$,   excites the odd
frequencies of the string, while the relative wave, ${\bf Z}(x^+)$,
 excites the
even frequencies. This, and the zero-mode contribution, 
are the only differences between these two types of wave. 
The average
square mass of the outgoing string, taken  initially 
in its ground state, is 
\eqn\avmass{  < M^2> \; =\; 4 {\bf w}_{\rm out}^2 + 
 \; 32\times \left(\; \sum_{n=1,3,  ..} \; 
\vert\;  {\bf y}(n/p^+)
\;\vert^{\; 2}
\;+\; 
 \sum_{n=2,4,..}\;   
\vert \; 
 {\bf  z}(n/p^+)\; \vert^{\; 2} 
\; \right)  \ .  
}
As can be easily seen,  the 
 excitation probabilities vanish 
 in the  limit $p^+\to 0$, provided  the shape of the background 
pulse is  smooth  (i.e. it  does not contain
arbitrarily high frequencies). This is the adiabatic limit, to
which we will now turn  our attention.


\newsec{Multipole Expansion}

  The results of the previous section make it clear that the most
relevant parameter in the problem is $\omega p^+ \alpha^\prime$,
where $\omega$ is the characteristic frequency of the pulse.
In general a  pulse will have many frequencies, possibly inside
different bandwiths. If  the maximal characteristic frequency is 
sufficiently small, 
\eqn\maxfr{\omega_{\rm max}\;  p^+ \alpha^\prime \ll 1 \ \ ,
}
the excitation probability will be practically zero. 
The string  will surf   in this case 
 on the wave,  and emerge  behind it having  suffered at most 
an adiabatic elongation or contraction. Of course,  when  $p^+=0$
the string  rides the wave forever, it is  indeed indistinguishable
from the massless strings that  {\it make} the wave.

We can   study the low-frequency  limit in more detail  with the help of 
 the Euler-MacLaurin formula, which  expresses the difference 
between a discrete sum  and the corresponding  integral~: 
\eqn\euler{  \sum_{n=0}^\infty\; \epsilon\; f(x-n\epsilon)\; = \;
 \int_{-\infty}^x  dy\;  f(y)\; + \; {\epsilon \over 2}\; f(x)\;
+ \; \sum_{r=1}^\infty\;  {B_{2r}\over 2r!}\;\epsilon^{2r}\; f^{(2r-1)}(x)\ .
}
Here $B_n$ are the Bernoulli numbers~,   $B_2= 1/6$, $B_4=-1/30\  \cdots$,
and $\epsilon$ a  small expansion parameter.  
 Using this formula we can express  the
function  ${\bf G}_{\rm in}(x/\lambda,\lambda)$,   given 
by the discrete sum  
 \specl\ , as an  asymptotic power series
in  $\lambda$ for fixed $x$~: 
\eqn\eulone{{\bf G}_{\rm in}(x/\lambda,\lambda) = 
{1\over \pi \lambda}  \int_{-\infty}^{\  x}  dy\; 
 {\bf \tilde Z}(y)\; + \; {1 \over 2}\; {\bf \tilde Z}( x)\; 
+ \; \sum_{r=1}^\infty\;  {B_{2r}\over 2r!}\; (\pi\lambda)^{2r-1}\;
 {\bf Z}^{(2r-1)}( x)\ ,
}
where
${\bf \tilde Z} \equiv {\bf Z} - \pi {\bf w}_{\rm in}$ as 
previously  defined. 
 The function ${\bf F}_{\rm in}(x/\lambda,\lambda)$ 
given  by the alternating sum  \speclc\ can
be expanded similarly. One  uses  the Euler-MacLaurin formula
for  the even and odd terms of the sum separately, and
then performs  a  Taylor expansion of their  difference. 
The first few  terms of the resulting series are~:
\eqn\eulf{ {\bf F}_{\rm in}(x/\lambda, \lambda) \; = \; 
{1\over 2}\;{\bf Y}(x)\; +\; {\pi\lambda \over 4}\;
 {\bf Y}^{\prime}(x)\; - \; 
{(\pi\lambda)^3\over 48}\; {\bf Y}^{\prime\prime\prime}(x)\; +\; \cdots
}
with primes denoting  derivatives of the function ${\bf Y}$. 
 All higher-order terms are of course easy to compute, if necessary.

 We want to  insert these   expansions in the equation 
\speccc\ for the string coordinates. It is  convenient to 
define  $\tilde \sigma \equiv  \pi/2-\sigma$, 
so that the center of the open string 
 is located at $\tilde\sigma =0$ . The arguments of 
${\bf F}_{\rm in}$ and ${\bf G}_{\rm in}$ in \speccc\ are
$\lambda = p^+$ and $x = x^+ -\tilde\epsilon$, 
where $\tilde\epsilon = p^+(\pi/2\mp \tilde\sigma)$. We need  therefore
to further expand \eulone\ and \eulf\ in powers of $\tilde\epsilon$. 
After some lengthy but straightforward algebra one  finds~: 
\eqn\monop{  \matrix{  
{\bf G}_{\rm in}(\tau-\sigma, p^+)\; -\; 
{\bf G}_{\rm in}(\tau+\sigma -\pi, p^+) \; = \; 
\cr  \cr  
\ \  \ \    \ =\; \left({\bf Z} 
 - \pi {\bf w}_{\rm in}\right)\; (2\tilde \sigma/ \pi)
 \; + \; {\bf Z}^{\;\prime\prime}\;
(p^+)^2\;  ({\tilde \sigma}^3/3\pi 
-\pi\tilde\sigma/12)\;  +
 \ o\;(\; {\bf Z}^{(4)}(p^+)^4\; ) \ \ , \cr} 
}
and 
\eqn\monopo{  \matrix{  
{\bf F}_{\rm in}(\tau-\sigma, p^+)\; +\; 
{\bf F}_{\rm in}(\tau+\sigma -\pi, p^+) \; = \; 
\cr  \cr  
\ \  \ \    \ =\; {\bf Y} 
 \; +  \;  {\bf Y}^{\prime\prime}\;
(p^+)^2\;  ({\tilde \sigma}^2/2
-\pi^2/8)\;  +
 \ o\;(\; {\bf Y}^{(4)}(p^+)^4\; ) \ \ . \cr} 
}
The functions  ${\bf Y}$ and 
${\bf Z}$ and all their  derivatives are evaluated 
in the above expressions at $x^+=p^+\tau$. 
Putting together everything,  we  arrive at the
following small-$p^+$ expansion of the transverse string coordinates, 
expressed as functions of the
lightcone time $x^+$ and of the string parameter $\tilde\sigma$~:  
\eqn\finalexp{ \delta {\bf X}_{\rm in}
  \ =\; {\bf Y}\;+ \;  \left({\bf Z} 
 - \pi {\bf w}_{\rm in}\right)\; {2\tilde \sigma\over \pi}\;+ \;
(p^+)^2\; ( {\bf Y}^{\prime\prime}\; + {2\tilde\sigma\over 3\pi}
{\bf Z}^{\prime\prime})  \left({{\tilde \sigma}^2\over 2}
-{\pi^2\over 8} \right)\; + \; o((p^+)^4)\;    \ .  
}

  As a  check of the above result, notice that it  obeys
the free worldsheet  equation,
$(\partial_\tau^2 -\partial_{\tilde\sigma}^2)\; \delta {\bf X}_{\rm in} = 0$, 
and  the  boundary conditions 
${\bf X}_{\rm in} + \delta {\bf X}_{\rm in}
= {\bf Y}\pm {\bf Z}$ at $\tilde\sigma = \pm \pi/2$. 
 Allowed  terms in
the   expansion respect,  furthermore,    two discrete symmetries~: 
(a)  invariance  under  orientation flip ($\tilde\sigma\to 
-\tilde\sigma$) combined with an exchange of the two D-branes
(${\bf Z} \to 
-{\bf Z}$) , and (b)  invariance under time reversal on  the worldsheet
($\tau\to  -\tau$) combined with the  spacetime reflection  $x^\pm\to
- x^\pm$. It is this second symmetry that explains the absence
of odd-derivative terms in \finalexp\ . 
We can in fact  use
these  discrete symmetries, together with the 
linearity of the response, the free-wave equation
and the boundary conditions, to give an alternative systematic derivation
of the multipole  expansion.

  Let us evaluate the series  \finalexp\  at a time $x^+$, 
long after the passage of the pulse. Since the derivatives of
${\bf Y}$ and ${\bf Z}$ go  to zero, 
 only the first two terms of the series   survive. They describe an open
string which   emerges   from the wave in its initial state of
oscillation, given  by  ${\bf X}_{\rm in}$. The linear term in 
$\delta {\bf X}_{\rm in}$ corresponds  to an 
 adiabatic elongation   or contraction, as  expected. 
The process of string excitation (or the  inverse process)  is not,
in other words, 
visible at any finite  order of the small-$p^+$  expansion.

  Before concluding this section, it is instructive to consider also
the  T-dual problem of an open  string in an electromagnetic wave. The 
classical  trajectory, eqs.    \tdone\ and \tdtwo , can be
expanded easily   with the help of 
 the Euler-Maclaurin formulae  \eulone\ and \eulf. The
 first few non-trivial terms are~:
 \eqn\finalexpp{ \delta {\bf X}^\prime_{\rm in}
  \ =\;{2}\int_{-\infty}^{\; x^+} {\bf B}(y)\; dy \; +\; p^+
 {\bf B}^\prime \left({{\tilde \sigma}^2}
-{\pi^2/ 4} \right)\; + \;
 p^+ {\bf A}^\prime \;\pi\tilde \sigma \; + o(\;(p^+)^5)\;    \ .  
}
The first term comes from the monopole 
 coupling to the gauge 
field ${\bf B}$, while the third term is due to the leading,  dipole 
coupling to  the gauge field ${\bf A}$. To check this, note that
a pair of  opposite unit charges have the following dipole coupling
to the electromagnetic field, 
$\delta {\cal L}  = 
F_{\mu\nu}\;  r^\mu \dot X^\nu + \cdots$, where $X$ and $r$ are the
center-of-mass and relative positions.  For a plane-fronted wave, and  in
 lightcone gauge,  this gives 
 $\delta {\cal L} =  - p^+ {\bf A}^\prime \cdot {\bf r}+ \cdots$ . 
If the charges are connected by a string, there is furthermore a
tension energy equal to \ 
$ {\bf r}\cdot  {\bf r} / (4\pi^2\alpha^\prime)$. The minimum
of the energy is thus at ${\bf r}\equiv  \delta {\bf X}^\prime_{\rm in}(\pi) -
\delta {\bf X}^\prime_{\rm in}(0)  
 = p^+ {\bf A}^\prime \pi^2$, in full agreement with the third
term in  \finalexpp.


\newsec{Shock waves,  Boomerangs and Other Beasts}

   The failure  of the multipole expansion to describe the process
of string
excitation is most evident in 
the limiting  case of a $\delta$-function pulse. Consider
a   pulse propagating  entirely
 on the left D-string in Figure 2. This corresponds to the choice
\eqn\deltone{ {\bf Y}(x^+)\; =\; {\bf Z}(x^+)  -\pi{\bf w}_{\rm in}
\; =\; {1\over 2}\; {\bf C }\;  \delta (x^+)\ , 
}
with  ${\bf C }$ an arbitrary  polarization  vector. 
We assume also, for definiteness, that  the open string is initially 
in its ground state and at rest, so that its energy equals tension
times length, $p^+ = p^- = \sqrt{2}\; \vert {\bf w}_{\rm in}\vert$.
Its classical worldsheet at later times is  given by 
the right-hand-side of equation \speccc. This 
 is a sum of $\delta$-functions, which in 
 the time interval\  
$N\pi p^+\leq x^+\leq (N+1)\pi p^+$\   is equal to~:
\eqn\deltwo{  \delta {\bf X}_{\rm in} = \cases{\ \ \ \   0\ \ \ 
\ \ \ \ \ \ \ \ \ \ \ \ \ \ \ \ \ \ \ \ \ \ \ \ \ \ \ \ \ \ \ \ \ \ 
\ \ \ \ \ \ \  {\rm for } \ \ N < 0   \cr
\ \  {\bf C }\;  \delta (x^+-p^+\sigma - Np^+\pi)\ \ \ \ \ \ \ \ 
\ \ \ \ \ \ {\rm for } \ \ N\geq 0 \ \ {\rm even}   \cr
 - {\bf C }\;  \delta (x^+-p^+(\pi- \sigma) - Np^+\pi)\ \ \ \ \ \  
{\rm for } \ \ N> 0 \ \ {\rm odd} \ \ .   \cr}\ 
}
This time-dependence 
can be   visualized in  Figure 3. The primary 
pulse propagates initially on  the
left D-string, and reaches the endpoint of the fundamental string
at time $x^+=0$. There  it sends a clone  ripple 
down the open string, while continuing  on its way unchanged. This sounds
like a violation of  energy, but recall that 
in the zero-string-coupling limit, in which we here work, the D-string
 tension is infinite and back reaction 
can thus be neglected. After the D-brane ripple has left the scene, 
its clone on the fundamental string
keeps bouncing back and forth, fliping  sign after every 
reflection. Clearly,  this behaviour cannot be  
captured by the multipole  expansion, which `predicts'  that the open string
must return to its ground state adiabatically, after the pulse has left. 

Two remarks are here in order. First, one may wonder how the  pulse
can be transmitted when the polarization vector ${\bf C}$ points in the
direction of the stretched  string. The answer is 
 that in the lightcone gauge the reparametrization invariance
is  completely fixed,

\bigskip
\ifig\moverot{Three snapshots of the stretched open string (a) before
being hit on the left by the  pulse, (b) immediately after, and (c) 
after the secondary  pulse  has been reflected once at the
right  endpoint. The string's
 recoil in the  direction of the pulse is not accurately depicted by 
this drawing. 
In the second-quantized theory, 
open and/or  closed strings will  be radiated away in the process.  
} {\epsfxsize4.0in\epsfbox{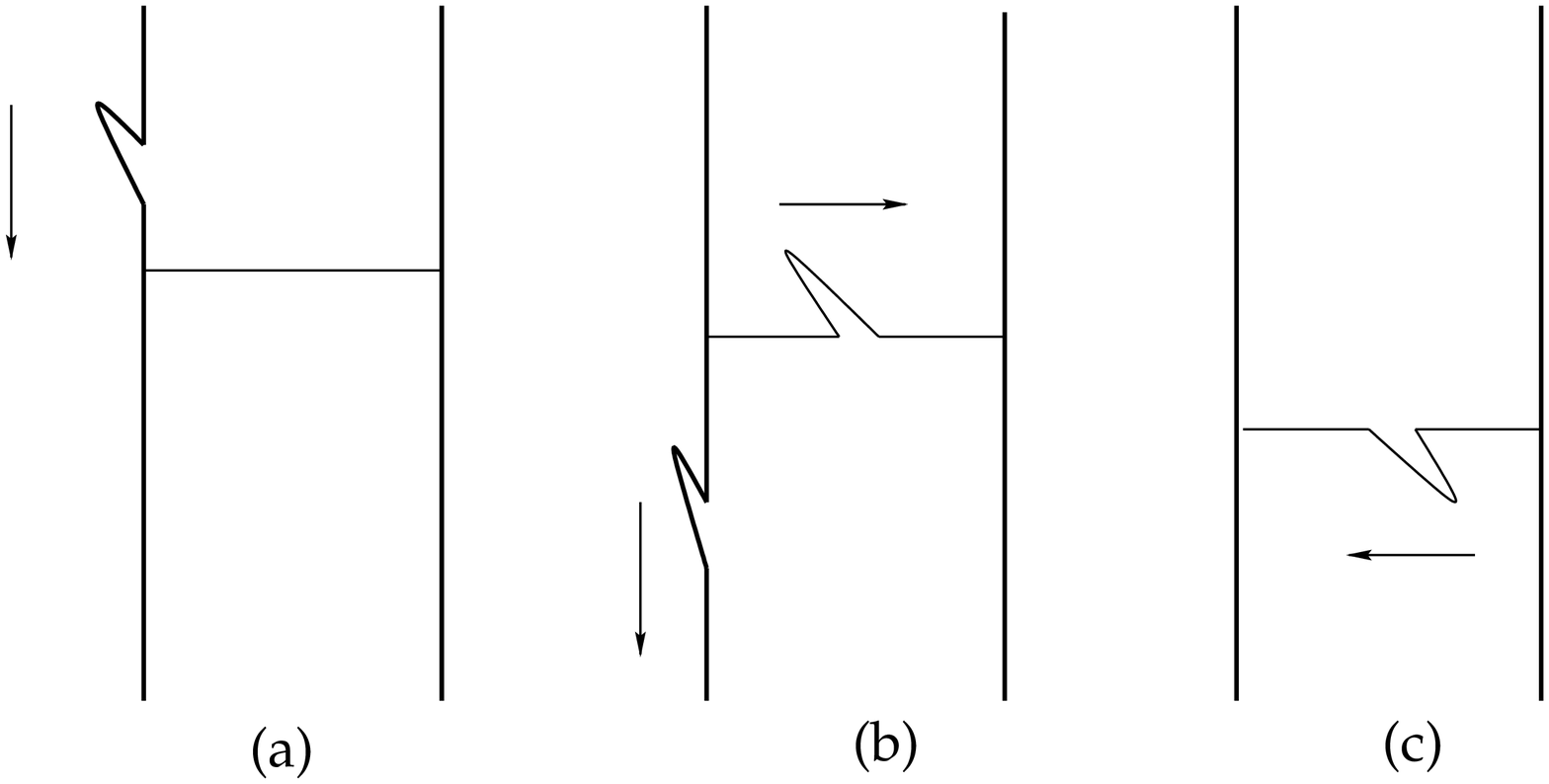}}

\noindent  so  longitudinal  oscillations
 are physical.
The second, more important  remark,  is that  for 
a  primary wave of arbitrary shape, the secondary ripples on the 
open string are given by superposing the  solutions \deltwo\ . This follows
from the linearity of the problem, and the fact that any temporal profile
is  a  superposition of  $\delta$-function pulses like \deltone . 
Thus, 
any  smooth  pulse of sufficiently short duration will 
give  the same qualitative behaviour as illustrated  in Figure 3. Pulses
of longer duration, on the other hand,  
will lead  to constructive or
destructive interference between the   disturbances
 entering the string from 
the left, and those   bouncing back and forth between its  
endpoints.

  It  should  be stressed  that although the transverse 
 coordinates ${\bf X}$ 
respond linearly to  the wave, the same  is not the case 
for the light-cone coordinate $X^-$. This is fixed  by the Virasoro
condition \lightc\ , which implies that $X^-$ is a quadratic
 function of the primary pulse. Note  in particular
that the string recoils after being  hit by the pulse, as can be deduced
from the expression for the 
center-of-mass momentum 
\eqn\recoil{ p^-(\tau)  = {1\over 2\pi p^+}\int_0^{\pi}  d\sigma\;
(\partial_\tau {\bf X}\cdot \partial_\tau {\bf X} 
+ \partial_\sigma {\bf X}\cdot \partial_\sigma {\bf X}) \ .    
}
The net recoil, $\delta p^-= p^-_{\rm out} -p^-_{\rm in}$,
 diverges for a $\delta$-function
pulse, and so  does the expectation value \avmass\  of the  mass squared 
operator in the quantum problem. 
This is not surprising, since  the  wave carries   infinite
energy and momentum in this limit. 
For a smooth but sufficiently short
pulse (for which  the interference terms can be neglected),  one 
finds  
\eqn\momtrsm{\delta p^-\; \simeq \;
 {2 T_F\over T_D}\;  P^-_{\rm pulse} \ ,   
}
where $T_F = (2\pi\alpha^\prime)^{-1}$ is the
tension of the fundamental string, and we have  used equation \enmo\ .  
Thus
the  outgoing  string  carries away  a fixed fraction
of the incident $p^-$ momentum in this limit. 
Note that since $T_D\sim 1/g_s$, this fraction is vanishingly small at
weak string coupling and the back reaction can be indeed neglected.

    The effects of an incident pulse with a  cusp (or kink)  can be
analyzed similarly 
with the help of equations \deltatwo\  and \avmass.   
If ${\bf Y}$,  or ${\bf Z}$,  behave  like $\vert x^+\vert^\beta$ near
the cusp, then their Fourier transforms ${\bf y}(\omega)$,  or 
${\bf z}(\omega)$,  vary   at high frequencies 
 like $\vert \omega\vert ^{-\beta}$. Since $\delta_2$ converges for
all (positive) values of $\beta$, the  string has a finite probability
to traverse  the pulse and remain 
in its ground state. The average mass squared of
the outgoing string  would, on the other hand, diverge
for   $\beta\leq 1/2$. This is  precisely the range of $\beta$ 
in which  the integral \enmo\  for 
the energy and momentum of the D-brane 
also diverges. We  conclude
that cusps are acceptable singularities, as long
as they  transport  finite energy and momentum.


  Having discussed the 
high-frequency components of the wave,
let me  next comment on  other potential sources of divergent behaviour.
One such source is of course a  highly monochromatic wave, which can produce
resonnant   excitation of the open string.
 As an example,  consider the following
helical D-string~:
\eqn\helix{ Y^2 + i Y^3\; =\; C\; e^{i\omega_0\; x^+}\;
\Longleftrightarrow \;
{\rm y}^2+i {\rm y}^3 \; =\; C\;
 \omega \; \delta (\omega -\omega_0)\ . 
}
The $\delta$-function  makes a divergent contribution to \avmass ,
whose origin is however easy to isolate. Indeed, using  
 Fermi's golden rule  for a wave train of total
duration $T$, and assuming that the incoming open string has a 
smooth  wavefunction $\psi(p^+)$ in momentum space
(normalized so that $\int dp^1 \vert \psi(p^+)\vert^2 = 1$)  we find~:
\eqn\fermi{   < M^2> \; = 
 \; {8 T  \over \pi}\;C^2\omega_0^2
  \times  \sum_{n\ {\rm odd}} \; 
\vert\;  { \psi}(n/\omega_0)
\;\vert^{\; 2}
  \ \ \ .  
}
Thus, for an incoming  open-string wavepacket, the divergence is   
regularized    by the  finite duration of  the pulse. Notice that
the energy drained out of the wave  
 by the  string grows only like the square root of time.

\bigskip
\ifig\moverot{Three configurations discussed in the main text~: (a)
regularized null D-brane scissors, (b) a boomerang, and (c) a
monochromatic  wave. 
} {\epsfxsize4.0in\epsfbox{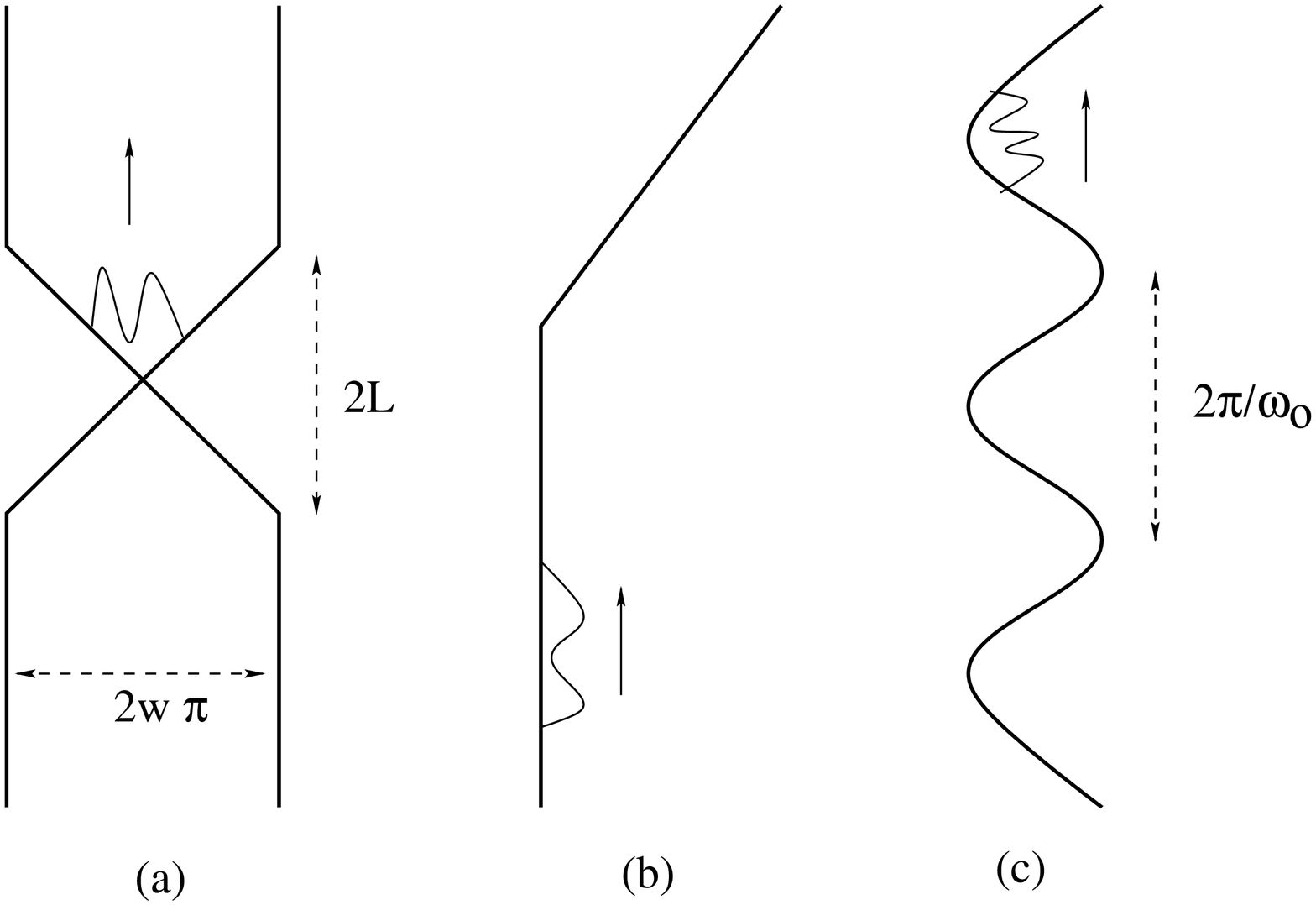}}

   A different  type of infrared problem  occurs for the null D-brane
scissors discussed in ref.   \bh. In this configuration ${\bf Z}$ grows
linearly with $x^+$, leading to an infinite adiabatic elongation
of the  stretched strings. As a result, 
 there are no normalizable states for  the
zero modes  except when $p^+=0$ \bh. To  isolate this  problem,  
 let us regularize   the large-time behaviour of the profile function. 
We can  choose, for instance, a piecewise-linear profile
like the one illustrated in
Figure 4~:
\eqn\nullsc{
{\bf Z}(x)\;  =\; \cases{ \  \pi {\bf w}\ \ \ \ \ \ \ \  
{\rm for} \ \ \  \ \   x\geq L \cr
\pi {\bf w} x/L     \ \ \ \ {\rm for} \ \ \  -L \leq x\leq L \cr
 - \pi {\bf w}\ \ \ \ \ \ \  {\rm for} \ \   \ \ \ x\leq -{L} \ \ .  \cr}
}
Its   Fourier transform 
can be  easily calculated  with the result~: 
\eqn\nullsf{ {\bf z}(\omega) = -i
 {\bf w}\; {{\rm sin}(\omega L)\over \omega L}\ .
}
The $ 1/\omega$ behaviour at high frequencies is characteristic
of the abrupt  kinks at $x=\pm L$. 
The basic  point, however, here  is that 
the sum  \avmass, which  is a measure of the string's  excitation, can
be made  arbitrarily small for large enough $L/p^+\alpha^\prime $. 
This shows that the null brane
intersection  does not, per se, introduce
any singularity, at least at  the  level of
first-quantized string theory. 

 The configuration  \nullsc\ is an open string analog of   
the parabolic orbifold `cosmology'  \hshs\lms, in which  the 
`universe' expansion  comes  to a stop. The 
null brane intersection 
plays the  role of  the  Big Bang/Big Crunch singularity, while
stretched open strings are 
analogous to twisted closed strings. The intersection point (like the
null orbifold singularity)  can be regularized by an extra  shift in 
one of the spectator dimensions. 
It was argued in ref. \bh\ 
that string collisions near the intersection point can  catalyze a
large backreaction, resulting  possibly in an intercommutation  
of the  D-strings. This would  allow any stretched open strings, 
formed during  the collision process, to escape with
finite energy to infinity. Indeed, after intercommutation 
each D-string has  a boomerang shape, 
${\rm Y}^2 (x^+) =  C
 x^+\theta(x^+)$, as illustrated in Figure 4b. An open string
moving on such a  D-brane will 
pick up  momentum  $p^2 = Cp^+$ while making the turn at $x^+=0$.   
The string's  energy in the outgoing  state will depend also
on its oscillation mass, but as the reader can easily verify,
this  stays  finite for finite $C$  (corresponding to a boomerang
making an  oblique angle). 

  A conceptual difficulty  from the
point of view of the open string, is that the process
of brane intercommutation looks,  superficially at least, non-causal.
The string only starts
stretching after it has crossed  the intersection point, by which
time it is too late to backreact on it. It would be very interesting to
see how the issue is resolved in the full
second-quantized string theory.

\newsec{From Lasers to  Gravitational Waves}

  The question we should finally address  is whether
 the coherent excitation of
strings,   discussed in this paper, can occur
 in a  real-world setting. 
There are   three kinds  
of relativistic strings in nature~:
QCD strings, fundamental strings,   and cosmic strings.\foot{To be sure,
the last two  kinds have yet to be experimentally detected! }
The fundamental and QCD strings are 
quantum mechanical, while cosmic strings
have  macroscopic sizes and are  to a good approximation classical. 
Strong coherent  radiation fields, on the other hand, 
are either electromagnetic or gravitational. As we have
seen,  the  necessary condition for string excitation is 
\eqn\exp{ \omega_0\;   p^+\  {\geq } \   T \ , 
}
where $\omega_0$  is the characteristic
frequency of the wave, and $T$ and $p^+$ are,  respectively,  
the  tension and lightcone momentum of the incident string. 
The first question is,  therefore~:   in what
physical settings can  condition  \exp\  be satisfied?

   The highest-frequency coherent electromagnetic fields are produced
by lasers,  which operate at or near  the  optical spectrum. Although coherent
stimulated emission in the soft X-ray region $\sim$ 10-15 nm
has been apparently reported (see for instance \lasers) most lasers
have wavelengths larger than a few hundred nanometers. 
This   corresponds  to frequencies 
$ \sim 4-6\; {\rm eV}$ at most. The Regge slope of QCD
strings, on the other hand, is   $\alpha^\prime =
(2\pi T)^{-1} \sim ({\rm GeV})^{-2}$. 
Coherent excitation of  hadrons  in a laser
beam is,  therefore,  possible for momenta 
$p^+ \geq 10^{8}\; {\rm GeV}$. This is beyond the reach of accelerators,
but within the upper end of the cosmic-ray spectrum. The flux of
cosmic rays at these energies is however exceedingly low,
roughly one  event per $m^2\times$ year \cosmic. Since the typical size
of a laser beam is in the $\mu$m   range, the rate of events
if one  placed  a laser in outer space  would still
be ridiculously small. Note, however, that since the  
 cosmic-ray  flux rises very rapidly with
decreasing energy (roughly by a factor of $10^3$ for each
energy decade)  a hard X-ray or $\gamma$-ray 
laser would fare much better!
 
The tension of fundamental open strings is at least six  orders
of magnitude above the tension of QCD strings, and probably 
 closer to the Planck scale.
This pushes the required momentum $p^+$ above $\sim 10^{14}$ GeV,
far above the GZK cutoff for  cosmic rays. 

  This brings us to  the third possibility, namely  cosmic strings \vil.
Their tension  in realistic particle-physics models
can range anywhere from the electroweak to the
grand-unification scale.   Strings formed during a phase
tansition in the early universe  stretch, on the other hand, 
over typical distances
$r_0$ of astronomical size. Since $p^+ \sim T r_0$,  the 
relevant criterion  for their  excitation  is  
\eqn\expone{ \omega_0\;  r_0\;   \geq\;   1\ . 
}
This puts essentially no restriction on  either electromagnetic
or gravitational waves.

 The  discussion in this paper was, to be sure,  about open strings
coupling to a plane wave through  their endpoints. Strictly speaking, 
this can be realized by `necklace' strings,  in which
magnetic monopoles play the role of the beads in a  necklace \vil. 
More generally, however, one may expect that if the wavefront  size
is much smaller than $r_0$, so that  the wave hits the string in 
a relatively localized region,  the results of this paper
could be carried over. If this were true, then a
 large cosmic string passing near 
the source of a strong gravitational wave,  could transmit a clone
of this gravitational  burst 
over an astronomically large distance (as in Figure 3).
Furthermore, since  propagation on the string is one-dimensional,
the amplitude of the clone wave  would stay  constant, while
the amplitude of the primary burst  in the bulk of space 
would  decrease  as the inverse of the distance from the source. This 
{teleportation}  of the  signal might  be relevant in  searches
for gravitational radiation.\foot{For a  review of gravitational wave
detection see \thorne. The string will ultimately release its energy
either by radiating gravitationally, or by splitting off small loops of
string -- see  for instance \vil\dam\ and references therein.
} The question is presently being  studied. 

  Let me finally point out that cosmic string networks are 
disfavoured as seeds of  large-scale anistropies, mainly because
their non-linear evolution tends to produce  incoherent 
primordial fluctuations  \acpeaks\acpeak\ruu.\foot{Though 
  current  CMB data is still, 
apparently,  compatible  with a sizable  component of 
cosmic-string seeds \acpeakss, and there is also scepticism
concerning the validity of the  calculations \ru.} 
It is worth investigating whether the simple model discussed here
could have any bearing on this issue. 


\vskip 1cm
{\bf Acknowledgements}: I thank  M. Blau, T. Damour, E. Floratos, 
J. Iliopoulos, P. Mora, G. Papadopoulos, B. Pioline, R. Russo and  
F.  Zwirner for useful conversations. 
This research was supported in part by  the European IHP Networks
`Superstring Theory' (HPRN-CT-2000-00122) and
`The Quantum Structure of Spacetime' (HPRN-CT-2000-00131).

\footatend\vfill\supereject\immediate\closeout\rfile\writestoppt
\baselineskip=14pt\centerline{{\bf References}}\bigskip{\frenchspacing%
\parindent=20pt\escapechar=` \input refs.tmp\vfill\eject}\nonfrenchspacing \bye